\documentclass[11pt]{article}
\usepackage{jcappub}
\usepackage{mathtools}
\usepackage{bm}
\usepackage{dsfont}
\usepackage{color}
\usepackage{array}
\usepackage{graphicx}
\usepackage{subcaption}
\usepackage{soul}
\usepackage{multirow}
\usepackage{multicol}
\usepackage{float}
\usepackage{hhline}
\usepackage[dvipsnames]{xcolor}
\usepackage[normalem]{ulem}
\usepackage{mathrsfs}
\usepackage{wasysym}
\usepackage[mathscr]{euscript}
\usepackage{ulem}

\title{Imprints of energy injection by compact dark stars in the 21-cm signal}

\author[a,b,c]{Boris Betancourt Kamenetskaia,}
\author[b]{Alejandro Ibarra,}
\author[d]{Chris Kouvaris}
\affiliation[a]{Cosmology, Gravity, and Astroparticle Physics Group, Center for Theoretical Physics of the Universe,
Institute for Basic Science (IBS), Daejeon, 34126, Korea}
\affiliation[b]{
Technical University of Munich,
TUM School of Natural Sciences, Physics Department,  James-Franck-Str. 1, 85748 Garching, Germany}
\affiliation[c]{\normalsize Max-Planck-Institut f\"ur Physik (Werner-Heisenberg-Institut), Boltzmannstr. 8, 85748 Garching, Germany}
\affiliation[d]{Physics Division, National Technical University of Athens, 15780 Zografou Campus, Athens, Greece}

\emailAdd{laybors@ibs.re.kr}
\emailAdd{ibarra@tum.de}
\emailAdd{kouvaris@mail.ntua.gr}

\abstract{A strongly self-interacting component of asymmetric dark matter particles can form compact dark stars. The high dark matter density in these objects may allow significant dark matter annihilation into Standard Model particles, even when the portals to the visible sector are extremely weak. In this paper we argue that compact dark stars could constitute an important source of energy injection during the cosmic dawn era in addition to that of the baryonic stars. Therefore, if dark stars annihilate into photons, the luminosity of dark stars may significantly raise the gas temperature of the Universe at small redshifts. This modification to the standard thermal history of standard Cosmology would have implications for the observed 21-cm signal and the process of reionization, thus providing a new probe for particle dark matter.
}

\begin{document}
\maketitle
\flushbottom

\section{Introduction}

Dark matter (DM) is a major component of the Universe, constituting approximately 27\% of its total energy content \cite{Planck:2018vyg}.
Yet, the nature of DM is still unknown, and its properties are  largely unconstrained. Up to now there is only evidence for this component through its gravitational effects on visible matter. On the other hand, all current efforts for detecting non-gravitational signals have not yield any evidence, indicating that the DM has very weak interactions with the Standard Model (SM) particles. However, DM particles could have sizable interactions with themselves or with other particles in the dark sector, which may produce observable signatures in collisions of galaxy clusters \cite{Randall:2008ppe} or on the shape, density profiles and substructure of DM halos, see {\it e.g.} \cite{1994Natur.370..629M,1995ApJ...447L..25B,2001AJ....122.2396D,Rocha:2012jg,Peter:2012jh,2015MNRAS.452.3650O,2022JCAP...07..031Z,2012MNRAS.422.1203B,2016A&A...591A..58P}.

One of the most important open questions in astroparticle physics is whether DM is symmetric or asymmetric in nature. An interesting possibility is that DM carry a charge under a global quantum number (akin to the baryon number), and that there is an asymmetry between DM particles and antiparticles (akin to the asymmetry between baryons and antibaryons in the visible sector). If DM has also strong self-interactions, part of the DM population could collapse, leading to formation of stable compact objects~\cite{2015PhRvD..92f3526K,2016JHEP...02..028E} (akin to the formation of stars in the visible sector). These compact dark stars (DSs) may be searched for in the sky using a variety of techniques, depending on their mass: stellar microlensing~\cite{Macho:2000nvd,EROS-2:2006ryy,Niikura:2019kqi}, supernovae magnification~\cite{Zumalacarregui:2017qqd}, gravitational waves from DS mergers~\cite{Maselli:2017vfi,Kavanagh:2018ggo,LIGOScientific:2019kan,Chen:2019irf}, dynamical constraints from dwarf galaxies~\cite{Brandt:2016aco} or wide binaries~\cite{Monroy-Rodriguez:2014ula}, or dwarf galaxy heating~\cite{Lu:2020bmd} (see~\cite{Green:2020jor} for a review of constraints). Based on the majority of the aforementioned data, it appears that a maximum of 1–10\% of the galaxy's DM could be composed of compact DSs, depending on the mass (and mass spectrum). 

DSs could also have other signatures. DM might  emit dark photons, which could convert via mixing with Standard Model photons that are potentially  detectable~\cite{Maselli:2019ubs}. Also, the compact DSs could  accrete protons and electrons from the interstellar medium, which would ultimately settle at the core of the DSs and thermalize with the compact DS. The hot gas of protons and electrons would then emit radiation, eventually producing very bright outbursts that can even reach up to $10^4$ solar luminosities~\citep{Betancourt_Kamenetskaia_2023}.  Finally, if the conserved  DM  quantum number is slightly broken, the DM particles could efficiently annihilate (or decay) into SM particles in the dense interior of the DSs, also producing potentially observable signatures in current gamma-ray or neutrino telescopes, even when the global symmetry is broken at the Planck scale ~\citep{Betancourt_Kamenetskaia_2023}. This can additionally cause a luminosity modulation if the DSs are subjected to radial oscillations triggered by the DM annihilation~\cite{Kouvaris:2025ijy}.  

In this paper we concentrate on this last possibility. DSs could have formed before ordinary stars, and the annihilation into photons at very early times could potentially produce a dramatic effect on the reionization of the Universe and the 21-cm spectrum.  Hints for the detection of a cosmological signal were first reported by the EDGES collaboration~\citep{2008ApJ...676....1B}, although their claim seems to be in tension with the results from SARAS~\citep{2013ExA....36..319P} and LEDA+\citep{2018MNRAS.478.4193P}. Future experiments, such as HERA~\cite{DeBoer:2016tnn} and REACH~\cite{deLeraAcedo:2022kiu} will have a much larger sensitivity to a cosmological 21-cm signal, and will provide invaluable information about possible new effects on the reionization of the Universe, like the one studied in this paper.

This work is organized as follows. In Section \ref{sec:formation} we revisit the process of formation of DSs and we argue that under certain conditions these could have formed before the first stars. 
In Section \ref{sec:E-injection} we estimate the rate of energy injection by DM annihilations in DSs, and in Section \ref{sec:signals} we discuss the impact of this energy injection on the form of the brightness temperature of the 21-cm line. Finally, in Section \ref{sec:conclusions} we discuss our results and present our conclusions. \footnote{Throughout the paper we will assume a spatially flat universe described by the $\Lambda$CDM model with  $H_0=100h\,{\rm km\,s^{-1}\,Mpc^{-1}}$, $h=0.696$, $\Omega_{m,0}=0.286$ and  $\Omega_{\Lambda,0}=0.714$ \citep{2016A&A...594A..13P}. We consider a power spectrum normalization with $\sigma_8=0.76$ at $z=0$, as computed using CAMB~\cite{Lewis:1999bs,Howlett:2012mh} with scalar amplitude $A_s=2.1\times10^{-9}$. We will also work in natural units $c=\hbar=k_B=1$.}

\section{Dark star formation}
\label{sec:formation}

Following~\citep{2019JCAP...03..036H}, we consider a dissipative model that consists of two particles: a dark electron, with mass $m_{e_D}$  constituting a fraction $f_{e_D}$ of the total DM abundance,  and a dark photon, with mass $m_{\gamma_D}$, with coupling strength to the dark electron parametrized by the structure constant $\alpha_D$. The parameters of our self-interacting DM model are  constrained by observations of halo shapes and the bullet cluster~\cite{Tulin:2017ara,Rocha:2012jg}. Concretely, if $f_{e_D}=1$ the self-scattering cross-section must satisfy
\begin{equation}
    \frac{\sigma_M}{m_{e_D}}\approx4\pi\frac{\alpha_D^2 m_{e_D}}{m_{\gamma_D}^4}\lesssim 1~\mathrm{cm}^2/\mathrm{g}.
    \label{eq:Moller-cross-section}
\end{equation}

DSs will form  by contraction and fragmentation of a primordial overdensity of dark electrons in the Universe. To trace the formation history of the DS, we consider an overdensity region of cold DM (CDM) (a proto-halo) that collapses concurrently with the dark electrons by the effect of gravity. The total mass of the CDM in the proto-halo is $M_{\rm halo}$, while the mass and density in the form of dark electrons is $M_{e_D}=f_{e_D}M_{\rm halo}$. To determine the densities in the proto-halo we use the spherical collapse model \cite{book_Mo_vdBosch_White}, in which overdensities decouple from the Hubble flow and ultimately ``turn around'' at a redshift $z_{ta}$, at which the density of the proto-halo is:
\begin{align}
\rho_{\rm DM}(z_{\rm ta})=\frac{9\pi^2}{16}\overline\rho_{\rm DM}(z_{\rm ta})
\end{align}
with $\overline\rho_{\rm DM}(z)$ the average DM density in the Universe at that redshift. Accordingly, the density of the dark-electron component of the proto-halo is $\rho_{e_D}(z_{ta})=f_{\rm e_D}\rho_{\rm DM}(z_{ta})$, and correspondingly the number density is $n_{e_D}=\rho_{e_D}/m_{e_D}$.
We will assume that at turn around the dark electron component has a temperature $T_{e_D}(z_{\rm ta})$. Furthermore, at some point during its evolution, the dark electron gas must become self-interacting so that it may collapse further than the rest of the CDM. This requires two conditions: i) the mean free path of the dark electrons must be smaller than the size of the collapsing CDM overdensity, and ii) the mean free path of the dark photons must be larger ({\it i.e.} the dark photons remain decoupled from the collapsing system of dark electrons and CDM). The first condition translates into an upper limit on the dark photon mass \citep{2019JCAP...03..036H}:
\begin{equation}\label{eq:m_gamma_D_chi_P}
    m_{\gamma_D}\lesssim 3.8\times10^8~\mathrm{eV}\left(\frac{1+z}{1+3400}\right)^{\frac12}\left(\frac{\alpha_D}{0.1}\right)^{\frac12}\left(\frac{f_{e_D}}{1}\right)^{\frac14}\left(\frac{m_{e_D}}{1~\rm GeV}\right)^{\frac14}\left(\frac{M_{\rm halo}}{10^{14}M_\odot}\right)^{\frac{1}{12}}.
\end{equation} Notice that this condition implies that not all DM can be in the form of dark electrons, since Eq.~(\ref{eq:Moller-cross-section}) along with Eq.~(\ref{eq:m_gamma_D_chi_P}) would require $M_{\rm halo}\gtrsim 10^{17}~M_\odot$ at $z\leq30$, which is at odds with the fact that the latest, most massive halos at the current cosmic time have masses of ${\cal O}(10^{15}~M_\odot)$. The second condition translates into a lower limit on the dark electron mass:
\begin{equation}
m_{e_D}\gtrsim3.7~\mathrm{MeV}\left(\frac{1+z}{1+30}\right)^{\frac23}\left(\frac{\alpha_D}{0.1}\right)^{\frac23}\left(\frac{f_{e_D}}{1}\right)^{\frac13}\left(\frac{M_{\rm halo}}{10^{10}M_\odot}\right)^{\frac19}.
\end{equation}
This requirement prevents tight coupling between dark electrons and photons. If the dark electron mass is too small, Compton scattering would render dark photons and electrons a single fluid, which would prevent energy loss and would halt DS formation.

To model the evolution with time of the dark electron clump we note that the conservation of energy implies that
\begin{equation}
\frac{dE}{dt}=-P_{e_D}\frac{dV}{dt}-\Lambda V,
\label{eq:dEdt}
\end{equation}
where $E$ is thermal energy of the dark electron gas enclosed in the volume $V$,  $P_{e_D}$ is the pressure of the gas of dark electrons with photon-induced self-interactions, which reads $P_{e_D}=n_{e_D}T_{e_D}+2\pi\alpha_Dn_{e_D}^2/m_{\gamma_D}^2$~\cite{2015PhRvD..92f3526K}, while $\Lambda$ is the rate of energy loss per unit volume. For dark electron bremsstrahlung, the total energy emission rate per unit volume in the non-relativistic regime is given by \citep{1975ZNatA..30.1546H,2019JCAP...03..036H}
\begin{equation}\label{eq:dark_bremss}
    \Lambda_{\rm \gamma_D}=\frac{32\alpha_{D}^3 n_{e_D}^2 T_{e_D}}{\sqrt{\pi}m_{e_D}^2} \sqrt{\frac{T_{ e_D}}{m_{e_D}}}\mathrm{e}^{-\frac{m_{\gamma_D}}{T_{e_D}}},
\end{equation}
where the exponential factor accounts for the suppression due to the nonzero mass of the dark photon when $m_{\gamma_D}\geq T_{e_D}$. The rate of energy loss per unit mass is 
\begin{equation}\Lambda=\Lambda_{\gamma_D}\, \mathrm{exp}\left[-R \sigma_{\rm C}n_{e_D}\right],
\label{lambda}
\end{equation}
where we have taken into account the fact that a fraction of the emitted dark photons would scatter with the dark electrons and would not escape the clump. Here, $R$ is the radius of the clump and $\sigma_{\rm C}=\frac{8\pi}{3}\frac{\alpha_D^2}{m_{e_D}^2}$ is the Compton cross-section for dark photons scattering off dark electrons.
Using that the energy of the homogeneous dark electron gas enclosed in volume $V$ is $E=(3/2) n_{e_D} V T_{e_D}$ and that the volume is related to the dark electron number density through $V=M_{e_D}/(n_{e_D}m_{e_D})$, Eq.~(\ref{eq:dEdt}) can be recast as:
\begin{equation}
    \frac{d\ln{T_{e_D}}}{d\ln{n_{e_D}}}=\frac{2}{3}\frac{P_{e_D}}{n_{e_D}T_{e_D}}-2\left[\frac{\Lambda}{3n_{e_D}T_{e_D}}\left(\frac{d\ln n_{e_D}}{dt}\right)^{-1}\right].
    \label{lnT}
\end{equation}

The evolution of the dark electron component can be separated into three stages. Since at the beginning of the evolution, the proto-halo is still very diluted, the interactions among dark electrons are rare and the energy loss via the bremsstrahlung of dark photons is practically negligible. Therefore, the collapse can be considered as adiabatic with pressure $P_{e_D}\approx n_{e_D}T_{e_D}$, and the evolution of the temperature of the dark electron gas is simply related to the dark electron density through:
\begin{equation}\label{eq:T_ff}
    T_{e_D}(z)=T_{\rm e_D}(z_{\rm ta})\left(\frac{n_{e_D}(z)}{n_{e_D}(z_{ta})}\right)^{\frac23}.
\end{equation}
This is the first stage of free-fall and this initial adiabatic collapse stops at the temperature and density $(T^J_{e_D},n^J_{e_D})$, when the mass of the dark electron component in the collapsing halo, $M_{e_D}$, becomes equal to its Jean mass, given by~\cite{2019JCAP...03..036H}
\begin{equation}\label{eq:const_jeans_mass}
     m_J=\frac{\pi}{6}\left(\frac{\pi}{G}\right)^{3/2}\frac{T^{3/2}_{e_D}}{m^2_{e_D} n_{e_D}^{1/2}} 
\end{equation}
with $T_{e_D}$ given in Eq.~(\ref{eq:T_ff}). After this epoch, the dark electron clump enters the second stage, where it continues losing energy via the bremsstrahlung of dark photons, re-virializing and contracting again, so that the total mass of the dark electron gas remains equal to the Jeans mass (contours of the Jeans mass are shown in the figure as gray lines). We will name this phase as the ``nearly virialized contraction'' (nvc) stage.  Using Eqs.~(\ref{eq:const_jeans_mass}) and (\ref{eq:T_ff}), the equality of the mass of the dark electron gas to the Jeans mass implies that the temperature $T_{e_D}$ of the dark electrons at the transition must be related to their density through:
\begin{align}
\label{eq:T_nvc}
    T_{e_D}=\left(\frac{6}{\pi}\right)^{\frac23}\left(\frac{G}{\pi}\right)m_{e_D}^{\frac43}f_{e_D}^{\frac23}M_{\rm halo}^{\frac23}{n_{e_D}}^{\frac13}.
\end{align}
In particular, this relation defines the location of the point $(T^J_{e_D},n^J_{e_D})$. In this phase 
$\frac{d\ln{T_{e_D}}}{d\ln{n_{e_D}}}=\frac13$, therefore using Eq.~(\ref{lnT}) and that the change in the internal energy of the gas is dominated by the energy loss via bremsstrahlung, one obtains that the dark electron density changes with the time as $\left(\frac{d\ln n_{e_D}}{dt}\right)^{-1}=\frac{n_{e_D}T_{e_D}}{2\Lambda}$.

Lastly, as the temperature and the density of the dark electron clump increase, so does the rate of dark photon bremsstrahlung emission. 
At large enough densities, the energy loss timescale becomes comparable to the so-called free-fall timescale, defined as  $t_{\rm ff}=\left(16\pi G \rho_{e_D}\right)^{-1/2}$~\cite{Penston:1969yy}. This marks the onset of the final phase named ``fragmentation'', which occurs at the temperature and density  $(T^{\rm frag}_{e_D},n^{\rm frag}_{e_D})$, determined by
\begin{equation}
\label{eq:free-fall_time}
   \frac{2\Lambda}{3n^{\rm frac}_{e_D}T^{\rm frac}_{e_D}}= \left(\frac{d\ln n_{e_D}}{dt}\right)\Big|_{n_{e_D}=n_{e_D}^{\rm frac}} = \left(16\pi G m_{e_D}n_{e_D}^{\rm frac}\right)^{1/2}.
\end{equation}
In this stage, the temperature of the dark electron gas depends on the density as $T_{e_D}\propto n_{e_D}^{-4/3}$, as follows from  eqs.~\eqref{lnT} and \eqref{eq:free-fall_time}, and the energy is rapidly evacuated as the clump continues collapsing, ultimately fragmenting to several self-gravitating chunks of matter with mass equal to the Jeans mass.

Eventually, the density of dark electrons becomes so large that the mean free path of dark photons becomes smaller than the size of the collapsing fragment, and the latter becomes optically thick. This occurs when the exponential term of Eq.~\eqref{lambda} starts to dominate and dark photons are reabsorbed before they can escape from the clump:
\begin{align}
R\sigma_c n_{e_D}>1.
\end{align}
After this moment, the fragmentation stops, and the energy is no longer evacuated from the bulk of the fragment but rather from the surface. This marks the formation of the minimal fragments, that is, the DSs.

\begin{figure}[t!]
	\centering
	\includegraphics[width=.6\textwidth]{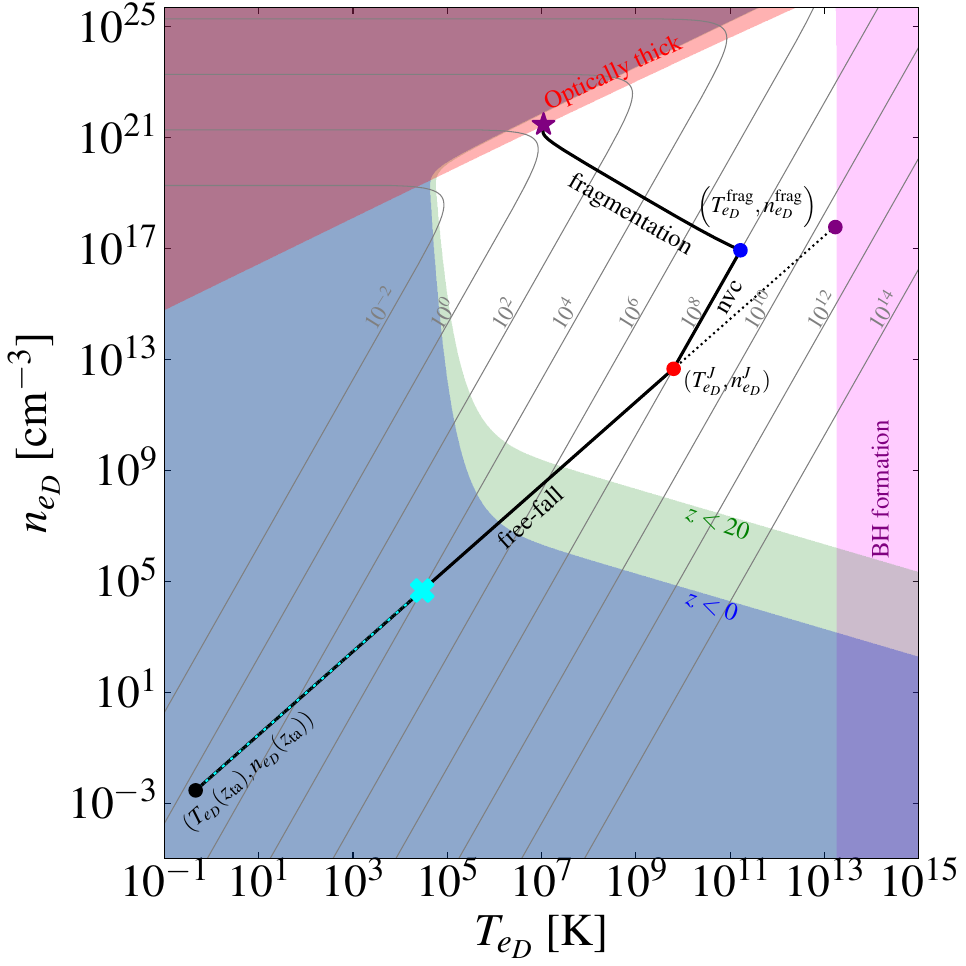}
	\caption{Sketch of the relation of the number density with the temperature of a dark electron clump in  
    a halo with $M_{\rm halo}=10^{5}M_\odot$ (dashed cyan line), $M_{\rm halo}=10^{9}M_\odot$ (black solid line) and $10^{13}M_\odot$  (dotted gray line), assuming that 10\% of the total mass of the halo is in the form of dark electrons. We take the parameter values: $m=10\ \rm GeV$, $m_{\gamma_D}=100\ \rm eV$ and $\alpha_D=0.1$. Solid gray curves: Contours of constant Jeans mass. Red-shaded region: Optical thickness region $R\sigma_c n_{e_D}>1$. Purple-shaded region: BH formation region. Green-shaded region: Region where the redshift of DS formation is less than 20. Blue-shaded region: Region where bremsstrahlung cooling is inefficient and DSs would not be formed today.}
	\label{fig:Formation_phase_space}
\end{figure}

In Fig.~\ref{fig:Formation_phase_space} we sketch the evolution of the dark electron clump number density and temperature. We consider the specific choice of parameters $m_{e_D}=10~\rm GeV$, $m_{\gamma_D}=100~\rm eV$, $\alpha_D=0.1$ and $f_{e_D}=10\%$. We present three different choices of the halo mass: for the solid black curve we consider $ M_{\rm halo}=10^{9} M_\odot$, while for the dashed cyan line we take $ M_{\rm halo}=10^{5} M_\odot$ and for the dotted gray line we set $ M_{\rm halo}=10^{13} M_\odot$. We fix the initial condition in all cases with the turnaround temperature and density represented by the black point. For our analysis we adopt $T_{e_D}(z_{\rm ta})=5\times10^{-3}T_{\rm CMB}(z_{ta})$ (while $T_{\rm CMB}(z_{\rm ta})\simeq 2.73(1+z_{ta})$), although our conclusions do not depend strongly on the specific value.
We focus first on the black solid curve, which represents a scenario of possible DS formation. Starting from the initial condition, the clump goes through the adiabatic free-fall stage until the Jeans mass becomes equal to the total dark electron mass $m_J=f_{e_D}M_{\rm halo}=10^8M_\odot$. This marks the transition to the nvc stage and is indicated in the figure as a red dot. We have chosen the turnaround redshift such that the red point is reached at $z=21$ (giving $z_{\rm ta}\sim33$). The nvc stage continues until the transition to the fragmentation phase, which is denoted by the blue point. Finally, fragmentation stops when the dark electron clump becomes optically thick and is presented as the purple star-shaped point in the figure.

It is important to note that not all DM halos will lead to the formation of DSs. Namely, if the halo is very massive, the contraction phase might lead to an extremely compact clump with a radius smaller than the Schwarzschild radius, thereby becoming a black hole. Concretely, this occurs when the radius of the clump with Jeans mass given by Eq.~(\ref{eq:const_jeans_mass}) becomes smaller than its Schwarzschild radius $R=2G m_J$. This requirement translates into $T_{e_D}\gtrsim T^{\rm sch}_{e_D}$ with 
\begin{equation}
 T^{\rm sch}_{e_D}= 6\times10^{12}~\mathrm{K}\left(\frac{m_{e_D}}{1~\rm GeV}\right).
\end{equation}
which is indicated in the figure as a purple band.
As apparent from the plot, a dark electron clump will end up as a black hole when $T^{\rm frag}_{e_D}\gtrsim T^{\rm sch}_{e_D}$, where the temperature at which the fragmentation starts can be calculated from the trajectory of constant Jeans mass, Eq.~(\ref{eq:T_nvc}), and the transition condition, Eq.~(\ref{eq:free-fall_time}): 
\begin{align}
    T_{e_D}^{\rm frag}&\approx \frac{9\pi^2 G}{64}\left[\frac{9\pi^3}{64}\left(\frac{\pi}{6}\right)^{\frac23}\right]^{-\frac34}\alpha_D^{-\frac32}m_{e_D}^{\frac52}f_{e_D}^{\frac12}M_{\rm halo},
\end{align}
where we have neglected the exponential factors in $\Lambda$. For illustration, we also show in Fig.~\ref{fig:Formation_phase_space}, as a dotted gray line, the relation between the density and temperature of a halo with mass $M_{\rm halo}=10^{13}M_\odot$. Clearly, the contraction of the halo leads to a black hole before starting the fragmentation phase. Furthermore, for very light halos the contraction and fragmentation phases might be slow on cosmological timescales and not be completed before $z\approx 6$, when reionization is finished, therefore causing no impact on the 21-cm signal.  This is again illustrated in Fig.~\ref{fig:Formation_phase_space}, as a dashed cyan line, which considers a halo with mass $M_{\rm halo}=10^5 M_\odot$. For this mass, the nvc phase is extremely slow, and is not even completed at redshift 0. This is denoted by the cyan x-shaped point, which lies completely in the blue-shaded region that shows the parameter space where the bremsstrahlung cooling is too inefficient to form DSs, such that the dark electron clump stays in the nvc phase to this day. 

From the three different trajectories in phase space shown in Fig.~\ref{fig:Formation_phase_space}, we understand that not all halos lead to DS formation. In fact, as we have discussed, if halos are too massive, the dark electron clump will collapse into a supermassive black hole without going through fragmentation. This leads to the definition of a maximum halo mass $M^{\rm max}_{\rm halo}$, above which no DSs are expected. We can estimate this mass from Eq.~\eqref{eq:T_nvc} and imposing $R=2Gm_J$ to be
\begin{equation}\label{eq:max_halo_mass}
    M^{\rm max}_{\rm halo}=1.1\times10^{15}M_\odot\left(\frac{\alpha_D}{0.1}\right)^3f_{e_D}^{-1}\left(\frac{m_{e_D}}{1~\rm GeV}\right)^{-3}.
\end{equation}

On the other hand, in the opposite case, if the halo is too light, bremsstrahlung cooling is too inefficient and the clump remains in the nvc phase to this day. This analogously suggests the definition of a minimum halo mass $M_{\rm halo}^{\rm m in}$, which we estimate from the implicit equation.
\begin{equation}\label{eq:M_halo_min_def}
    \left(\frac{d\ln n_{e_D}}{dt}\right)^{-1}\Big|_{\rm nvc}=\frac{n^J_{e_D}(M^{\rm min}_{\rm halo})T^J_{e_D}(M^{\rm min}_{\rm halo})}{2\Lambda(M_{\rm halo}^{\rm min})}=(t_U-t_{\rm nvc}(z)).
\end{equation}
Here $t_U\sim13.8~\rm Gyr$ is the current age of the Universe and $t_{\rm nvc}$ is defined as the age of the Universe when the clump reached the nvc phase ate redshift $z$. Therefore, the quantity $(t_U-t_{\rm nvc}(z))$ can be understood as the time left between the start of the nvc phase and today. The central part of Eq.~\eqref{eq:M_halo_min_def} can be understood as the timescale of formation of DSs starting from the nvc phase, hence Eq.~\eqref{eq:M_halo_min_def} imposes that the DSs have enough time to be formed by the present day. 

\begin{figure}[t!]
	\centering
	\includegraphics[width=.6\textwidth]{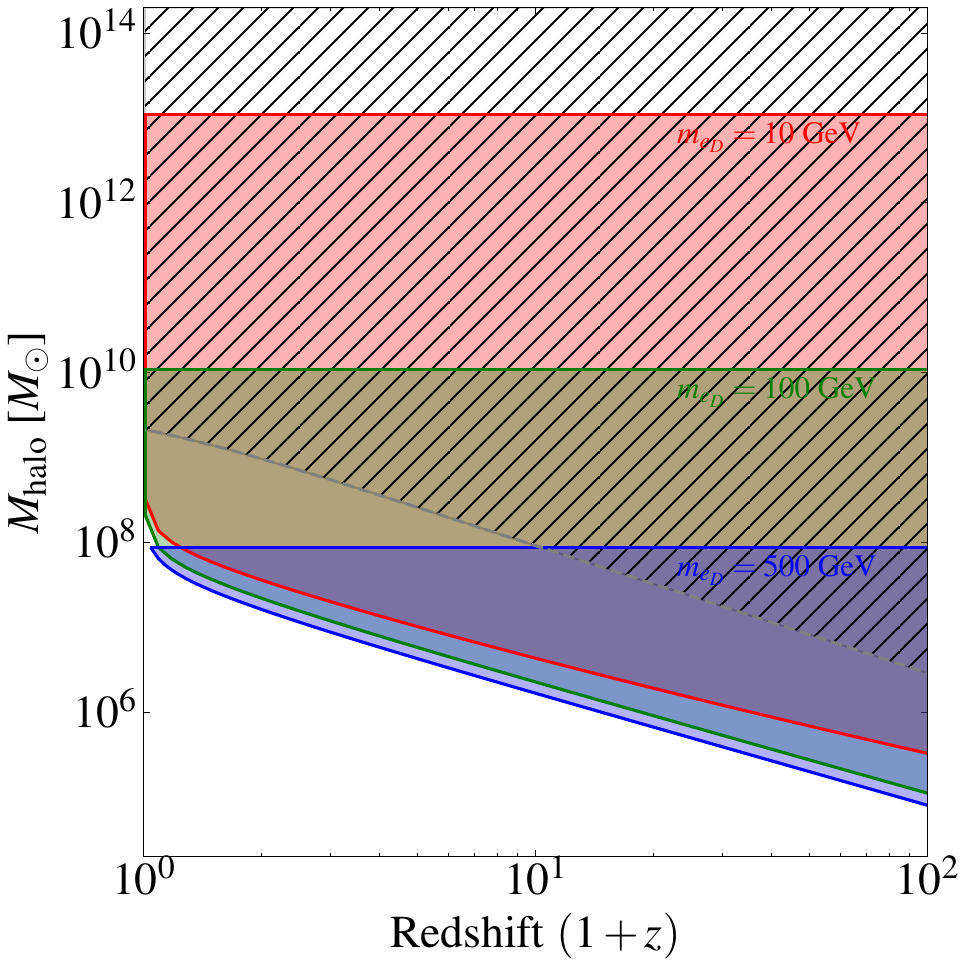}
	\caption{Range of halo masses leading to DS formation at redshift $z$, for $m_{e_D}=10$, 100 and 500 GeV. The rest of parameters is as in Fig.~\ref{fig:Formation_phase_space}. The hashed region corresponds to the halo masses that lead to baryonic star formation.}
	\label{fig:M_Halo_Min}
\end{figure}

We show in Fig.~\ref{fig:M_Halo_Min} the range of halo masses at a given redshift that lead to the production of DSs, 
for $m_{e_D}=10$ GeV (red-colored region), $100$ GeV (green-colored region) and $500$ GeV (blue-colored region), keeping the remaining parameters as in Fig.~\ref{fig:Formation_phase_space}. All halos with masses in this range will lead to a population of DSs at redshift $z$. The hashed region corresponds to the halo masses that lead to baryonic star formation with the respective $M_{\rm halo}^{\rm min}$ found by solving Eq.~\ref{eq:T_virial} for $T_{\rm vir}=10^4~\rm K$ in appendix~\ref{app:direct}.

\begin{figure}[t!]
	\centering
	\includegraphics[width=.5\textwidth]{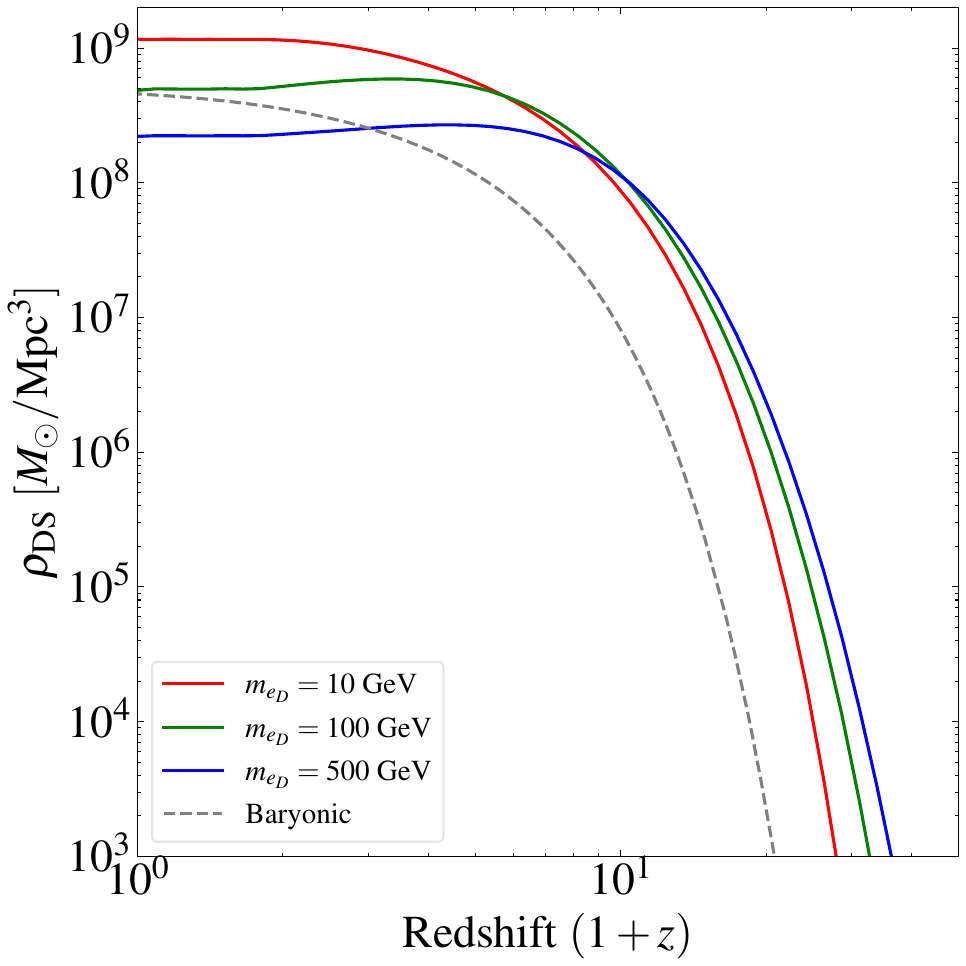}
	\caption{Star mass density as a function of redshift. The colored curves correspond to the benchmarks from Fig.~\ref{fig:M_Halo_Min} and the black dotted curve is for baryonic stars.}
	\label{fig:DSMD}
\end{figure}

The DS mass density  at redshift $z$  (or alternatively the total mass in form of DSs per comoving volume) can be estimated as:
\begin{align}
    \rho_{\rm DS}(z)=f_{\rm DS}f_{e_D}\frac{\Omega_{\rm DM,0}}{\Omega_{\rm m,0}}
    \int_{M_{\rm halo}^{\rm min}(z)}^{M_{\rm halo}^{\rm max}(z)}{M\,\frac{dn}{dM}(M)\,dM}.
\end{align}
where $\frac{dn}{dM}(M) dM$ is the comoving number density of haloes with masses between $M$ and $M+dM$, $\Omega_{\rm DM,0}/\Omega_{\rm m,0}\simeq 5$ is the ratio of the DM and visible matter density parameters today (and any cosmic epoch), $f_{e_D}$ is the fraction of DM in the form of dark electrons, and $f_{\rm DS}$ is the fraction of dark electrons in the halo that are in the form of DSs (the rest constitutes a diffuse component). In this paper we will assume $f_{*,\rm DS}=0.5$, in analogy to the measured fraction of visible matter in the form of stars.  We will model the number density of halos following Press \& Schechter \citep{1974ApJ...187..425P}. In this case, the total mass density of the collapsed haloes at redshift $z$ is:
\begin{equation}\label{eq:f_coll_DS}
     \int_{M_{\rm halo}^{\rm min}(z)}^{M_{\rm halo}^{\rm max}(z)}{M\,\frac{dn}{dM}\,dM}=\rho_{\rm m}(z)\left[\mathrm{erfc}\left(\frac{\delta_c(z)}{\sqrt{2}\sigma(M_{\rm halo}^{\rm min}(z))}\right)-\mathrm{erfc}\left(\frac{\delta_c(z)}{\sqrt{2}\sigma(M_{\rm halo}^{\rm max}(z))}\right)\right],
\end{equation}
where $\rho_{m}(z)$ is the total matter density, $\delta_c(z)$ is the linear overdensity at virialization and $\sigma^2(M)$ is the variance of the density field when smoothed on scale $M$, and which are given in \citep{2013A&C.....3...23M}. Fig.~\ref{fig:DSMD} shows the DS mass density as a function of the redshift for the considered benchmark values, as well as the baryonic case for comparison.
A pertinent observation from Fig.~\ref{fig:DSMD} concerns the DS scenario: the mass density reaches a peak before $z=0$, after which it either remains constant (e.g., for $m_{e_D} = 10~\mathrm{GeV}$) or declines slightly (e.g., for $m_{e_D} = 100,~ 500~\mathrm{GeV}$). This behavior contrasts with the baryonic case, where the mass density continues to increase. The difference arises from the presence of the upper mass limit, $M_{\mathrm{halo}}^{\mathrm{max}}$, in the DS model. At low redshifts, more massive halos, which are more likely to exceed $M_{\mathrm{halo}}^{\mathrm{max}}$, become increasingly common, thereby suppressing further growth in the DS mass density.

\section{Energy injection by dark stars}\label{sec:E-injection}

So far we have assumed the existence of a ``dark matter number" (akin to the electric charge or the baryon number) that prevents the decay or the annihilation of DM particles into Standard Model particles. If DM is of asymmetric nature,  observations indicate that the rate for  DM number violating processes must be tiny, so that large amounts of DM are still present today in galaxies and in the Universe at large. However, observations do not require the absolute conservation of the DM number. In particular, DM particles in the dense medium of a DS could be annihilating into Standard Model particles, thereby providing a test of the conservation of the DM number. 

The annihilation rate into Standard Model particles reads
\begin{align}
\Gamma_{\rm ann}=\frac{1}{2}\int dV \left(\frac{\rho(r)}{m_{e_D}}\right)^2 \langle\sigma v\rangle_{\rm ann},
\label{eq:ann-rate}
\end{align}
where $\langle\sigma v\rangle_{\rm ann}$ in the annihilation cross-section and $\rho(r)$ is the density distribution inside the DS. To calculate the density profile inside the DS we use the Tolman–Oppenheimer–Volkoff (TOV) equation, which describes the pressure distribution $P(r)$
\begin{subequations}\label{eq:DSStellar Structure}
	\begin{align}
		&\frac{dP}{dr}=-\frac{G M \rho}{r^2}\left(1+\frac{P}{\rho}\right)\left(1+\frac{4\pi r^3 P}{M }\right)\left(1-\frac{2GM}{r}\right)^{-1}.
	\end{align}
\end{subequations}
Here, $M(r)$ is the mass enclosed in the sphere of radius $r$, which can be obtained from the mass equation
\begin{eqnarray}
	 \frac{dM}{dr}=4\pi r^2\rho, \label{DSStellarStructureA} 
\end{eqnarray}
where the density distribution is related to the pressure $P(r)$ through the equation of state.
For a gas of dark electrons interacting via a dark photon, the equation of state is given in the following parametric form ~\citep{2015PhRvD..92f3526K}
\begin{subequations}\label{eq:EOS}
	\begin{align}
		&\rho(x)=m_{e_D}^4\left[\xi(x)+\frac{2\alpha_D}{9\pi^3}\left(\frac{m_{e_D}}{m_{\gamma_D}}\right)^2x^6\right], \\
		& P(x)=m_{e_D}^4\left[\psi(x)+\frac{2\alpha_D}{9\pi^3}\left(\frac{m_{e_D}}{m_{\gamma_D}}\right)^2x^6\right],
	\end{align}
\end{subequations}
where the functions $\xi$ and $\psi$ are given by
\begin{subequations}
	\begin{align}
		& \xi(x)=\frac{1}{8\pi^2}\left[x\sqrt{1+x^2}(1+2x^2)-\ln\left(x+\sqrt{1+x^2}\right)\right], \\
		& \psi(x)=\frac{1}{8\pi^2}\left[x\sqrt{1+x^2}(2x^2/3-1)+\ln\left(x+\sqrt{1+x^2}\right)\right].
	\end{align}
\end{subequations}
We solve the TOV and mass equations using the equation of state above with the boundary conditions $M(0)=0$, namely imposing that the mass distribution is not singular at the origin, and that $P(R_{\rm DS})=0$, namely that the pressure at the boundary of the DS vanishes. The boundary of the DS is given by its radius, $R_{\rm DS}$, which is implicitly defined by $M(R_{\rm DS})=M_{\rm DS}$, with $M_{\rm DS}$ the total mass of the DS. Finally, after determining the pressure distribution with the TOV equation, we determine the density distribution using the equation of state. 

\begin{figure}[t!]
	\centering
	\includegraphics[width=.48\textwidth]{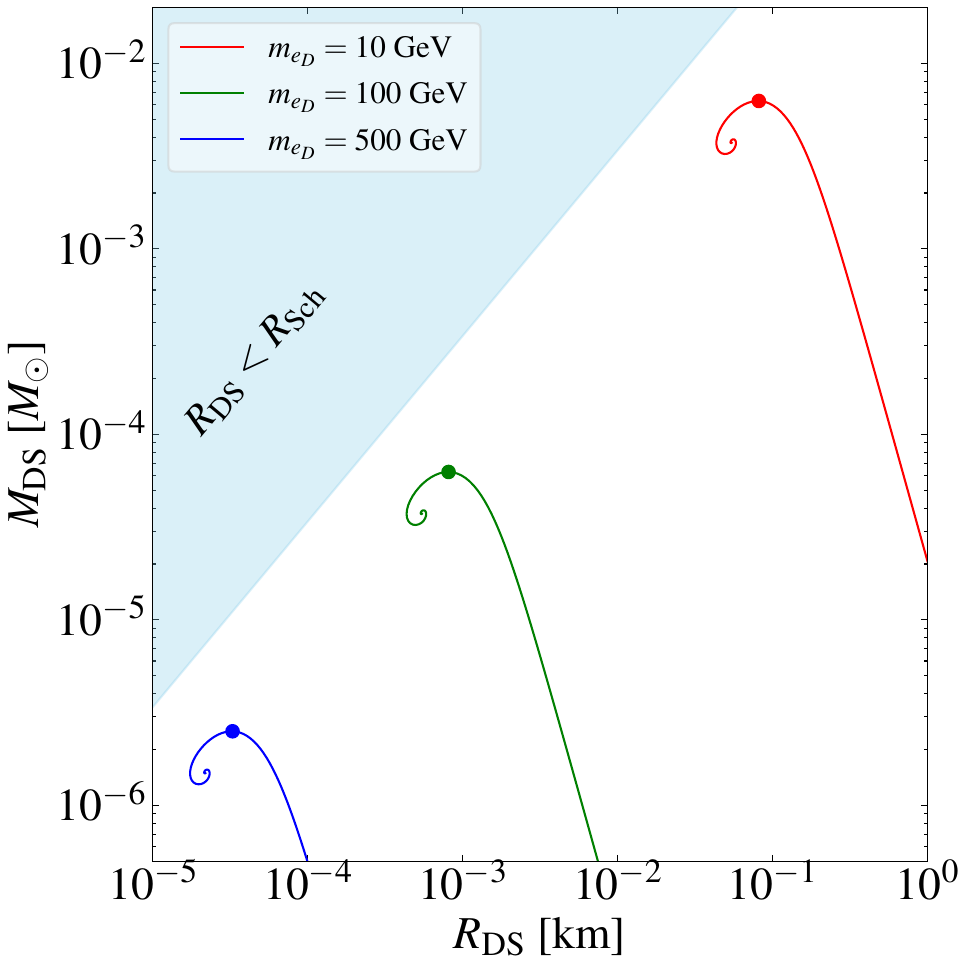}
	\includegraphics[width=.48\textwidth]{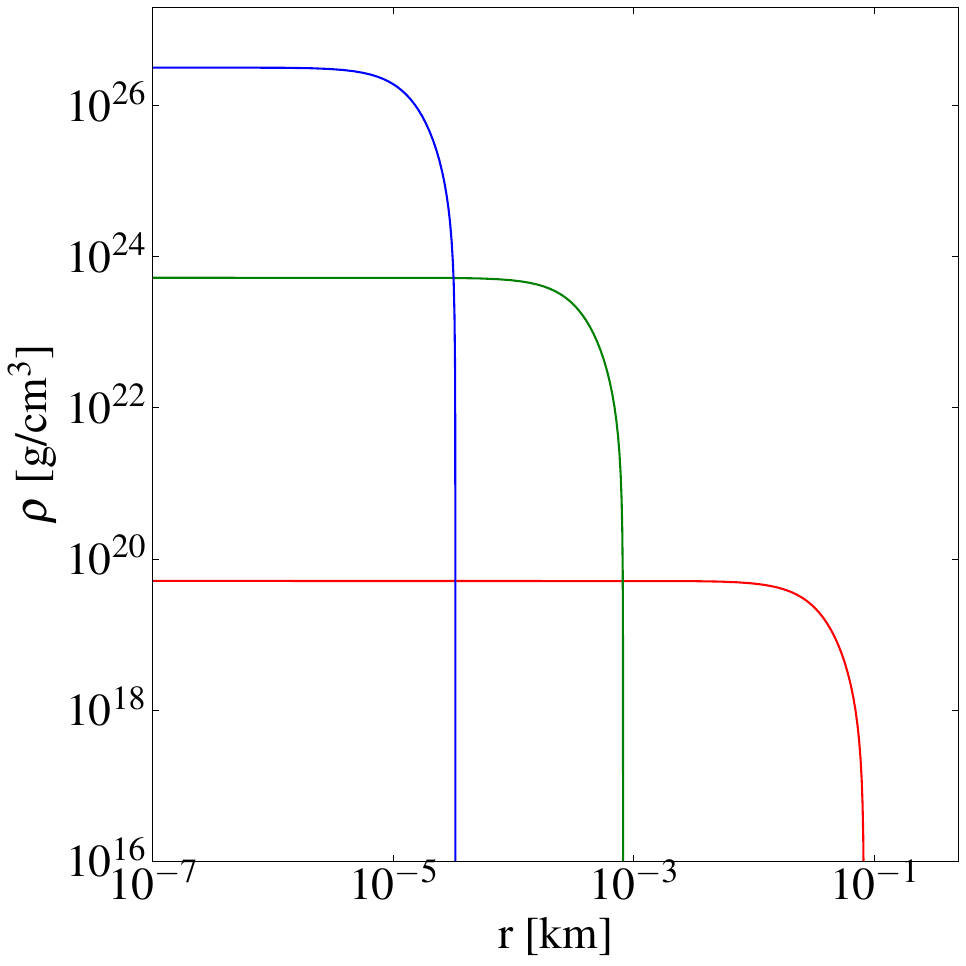} \\
	\includegraphics[width=.48\textwidth]{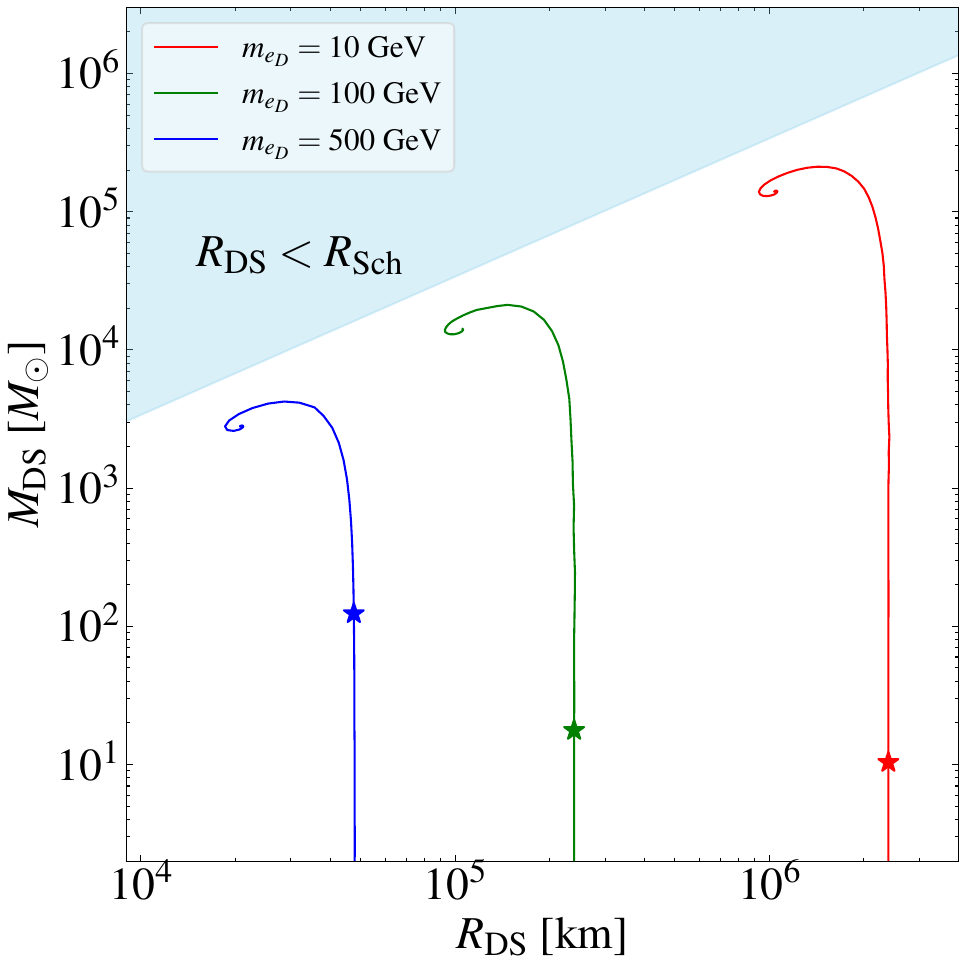}
	\includegraphics[width=.48\textwidth]{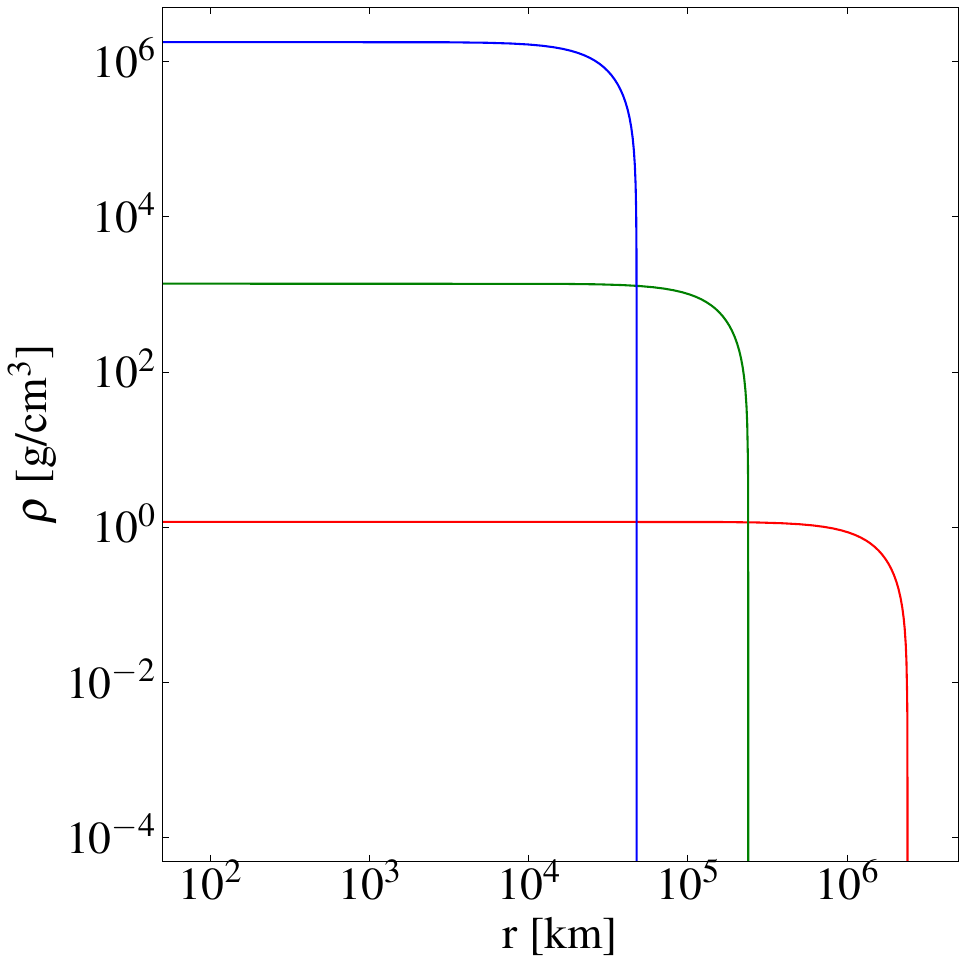}
	\caption{Mass-Radius relation (left panels) and representative density distribution (right panels), for a DS constituted by dark electrons without self-interactions (upper panels) and with self-interactions mediated by dark photons with mass $m_{\gamma_D}=100$ eV and coupling $\alpha_D=0.1$ (lower panels). The density distributions correspond to the benchmark points indicated in the left plots. }
	\label{fig:mass_vs_r}
\end{figure}

In the left panels of  Fig. \ref{fig:mass_vs_r}, we show the mass-radius relation of asymmetric DSs for a dark electron mass equal to 10 GeV (red), 100 GeV (green) and 500 GeV (blue), for the case where DM does not have self-interactions (upper panel, $\alpha_D=0$), or the strength of the self-interaction is non-zero (lower panel, $\alpha_D=0.1$, $m_{\gamma_D}=100$ eV). Each of the mass-radius curves has a maximum value for the asymmetric DS mass, in analogy to the ``Chandrasekhar mass" of white dwarfs, and only the branch of the curve with increasing radius from the maximum leads to stable configurations.  The light-blue colored area denotes the region where the radius is smaller than the Schwarzschild radius. In the right panels we show the density distribution inside the DS for the three benchmark points indicated in the left panels. As apparent from the plots, in the self-interacting case the DS is larger, heavier and less dense than in the non-self-interacting case. This is because the repulsive interaction we have assumed provides sufficient pressure to support the star in contrast to the non-interacting case where pressure is provided by degeneracy (Fermi pressure) which explains the large central densities required to support the star. Nonetheless, in both cases it is noteworthy the very large DM densities at the core of the DS, which therefore offer a unique environment to probe the conservation of the DM number. We point out that for Figs.~\ref{fig:Formation_phase_space}-\ref{fig:DSMD} in Sec.~\ref{sec:formation}, we have assumed that the self-interaction is repulsive throughout all stages in the evolution of the dark electron density. An attractive interaction is not required for the formation of dark stars. During the contraction and fragmentation phases, bremsstrahlung cooling is the dominant effect, while the self-interaction becomes important once the dark stars have formed.
Once we have determined the density inside the DS, one can calculate from Eq.~(\ref{eq:ann-rate}) the DM annihilation rate. We obtain
\begin{align}
\Gamma_{\rm ann}&=2\pi\langle\sigma v\rangle_{\rm ann} R_{\rm DS}^3 \left(\frac{\rho_c}{m_{e_D}}\right)^2 J_\star \cr
&\simeq \left(2.0\times10^{34}\ \mathrm{s}^{-1}\right)\left(\frac{\langle\sigma v\rangle_{\rm ann}}{10^{-44}\ \mathrm{cm}^3\, \mathrm{s}^{-1}}\right)\left(\frac{R_{\rm DS}}{10^5\ \rm km}\right)^3\left(\frac{\rho_c}{10^{3}\ \mathrm{g}\,\mathrm{cm}^{-3}}\right)^2\cr
&\times\left(\frac{m_{e_D}}{100\ \mathrm{GeV}}\right)^{-2}\left(\frac{J_\star}{10^{-2}}\right),
\end{align}
where we have defined $J_\star\equiv\int_{0}^{1}{d\hat{x}\, \hat{x}^2 \hat{\rho}^2(\hat{x})}$ as the dimensionless ``J-factor'', which encompasses the difference on the choice of the particular DS solution. Here, we have defined $\hat{x}\equiv r/R_{\rm DS}$ and $\hat{\rho}=\rho/\rho_c$ with $\rho_c$ the DS's central density. For the benchmark points $m_{e_D}=$ 10 GeV, 100 GeV and 500 GeV we find $J_*\approx0.012$ for the non-interacting case and $J_*\approx0.051$ for the interacting example with $m_{\gamma_D}=100$ eV and $\alpha_D=0.1$. 
Finally, the luminosity in gamma-rays generated by annihilations in the DS reads
\begin{align}
	L_{\rm ann}&=2m_{e_D} f^\gamma_{\rm ann}\Gamma_{\rm ann}\cr
	&\simeq  1.6 L_\odot f^\gamma_{\rm ann}\left(\frac{\langle\sigma v\rangle_{\rm ann}}{10^{-44}\ \mathrm{cm}^3\, \mathrm{s}^{-1}}\right)\left(\frac{R_{\rm DS}}{10^5\ \rm km}\right)^3\left(\frac{\rho_c}{10^{3}\ \mathrm{g}\,\mathrm{cm}^{-3}}\right)^2\cr
	&\times\left(\frac{m_{e_D}}{100\ \mathrm{GeV}}\right)^{-1}\left(\frac{J_\star}{10^{-2}}\right),
\end{align}
where $f^\gamma_{\rm ann}$ is the fraction of the total energy that goes into photons.
Here we have calculated the rate of energy injected by a single DS into its surroundings. We can now calculate the total rate of energy emitted by the cosmological DS population per comoving volume, the so-called luminosity density or in our case, the energy injection term
\begin{equation}\label{eq:energy_inj}
    \left(\frac{dE}{dVdt}\right)_{\rm inj}=L_{\rm ann}\left(\frac{\mathrm{\rho_{\rm DS}(z)}}{M_{\rm DS}}\right),
\end{equation}
where we have used the monochromatic approximation, in which we assume all DSs have the same mass $M_{\rm DS}$.

\section{Signals of annihilations inside dark stars in the intergalactic medium}
\label{sec:signals}

The DS population will inject energy into the intergalactic medium (IGM) through collisions with the intergalactic gas, increasing its temperature and enhancing the fraction of ionized atoms. This transfer of energy will lead to observable effects on the brightness temperature, which measures the difference between the CMB temperature and the temperature of a hydrogen patch at a given redshift. This quantity, neglecting spatial inhomogeneities, is given by \citep{2005ApJ...626....1B,2012RPPh...75h6901P}
\begin{align}
    \delta T_b&\approx\frac{T_s-T_{\rm CMB}}{1+z}\tau,
    \label{deltaT}
\end{align}
where $z$ is the redshift, $T_s$ is the so-called spin temperature, which characterizes the ratio between the occupation number of the two hyperfine levels of the neutral hydrogen atom (separated by an energy of $5.87\mu$eV, requiring a photon with a wavelength of 21 cm for the excitation), and $\tau$ is the optical depth for resonant 21-cm absorption, given by
\begin{equation}
    \tau=\frac{32 \lambda_{\rm 21\ \rm cm}^2 A_{10} n_{\rm H}}{16 T_s H(z)}.
\end{equation}
Here, $n_{\rm H}$ is the number density of neutral hydrogen and $A_{10}\approx2.9\times10^{-15}\ \mathrm{s}^{-1}$ is the Einstein coefficient, that describes the spontaneous emission rate of radiation \cite{1986rpa..book.....R,2019cosm.book.....M}.
Numerically,
\begin{align}
    \delta T_b
    &\approx(27\ \mathrm{mK})\, (1-x_e) \left(\frac{1+z}{10}\right)^{\frac12}\left(\frac{T_s-T_{\rm CMB}}{T_s}\right),
\end{align}
where $x_e$ is the fraction of free electrons.
The time evolution of the spin temperature is mainly determined by three processes: {\it i}) the absorption and emission of CMB photons by cosmic hydrogen; {\it ii}) spin-flip transitions due to collisions wtih other particles; and {\it iii}) spin-flip transitions induced by scatterings of  Ly$\alpha$ photons from the first stars ({\it i.e.} population III and II stars), known as the Wouthuysen-Field effect \citep{1952AJ.....57R..31W, 1958PIRE...46..240F}. Assuming that the rate of these processes is fast relative to the line  deexcitation time, the spin temperature can be well approximated by~\citep{1958PIRE...46..240F}:
    \begin{equation}\label{eq:spin_temperature}
        T_s^{-1}\simeq \frac{T_{\rm CMB}^{-1}+x_\alpha T_\alpha^{-1}+x_{\rm coll} T_{\rm gas}^{-1}}{1+x_\alpha+x_{\rm coll}},
    \end{equation}
where $x_\alpha$ is the Ly$\alpha$ coupling to the spin temperature, $x_{\rm coll}$ is the collisional coupling, $T_\alpha$ is the temperature of the Ly$\alpha$ radiation field at the Ly$\alpha$ frequency, and $T_{\rm gas}$ is the gas temperature. 
More concretely, the Ly$\alpha$ coupling to the spin temperature is given by \citep{2006MNRAS.372.1093F,2007MNRAS.376.1680P}
    \begin{equation}\label{eq:Ly_alpha_coup}
        x_{\alpha}=\frac{16\pi^2T_\star e^2 f_\alpha}{27 A_{10} T_{\rm CMB} m_e}S_{\alpha}J_\alpha,
    \end{equation}
where $f_\alpha\simeq 0.42$ is the oscillator strength of the Ly$\alpha$ transition, $T_\star=2\pi/\lambda_{\rm 21\ \rm cm}=0.0681\,\rm K$ is the 21-cm transition temperature, $e$ is the electron charge, $m_e$ is the electron mass,  $J_\alpha$ is the total Ly$\alpha$ background flux and $S_\alpha$ is a correction factor. The expressions for $J_\alpha$ and $S_\alpha$ are complicated and can be found in appendix \ref{sec:app_lyalpha}. The temperature $T_\alpha$ is well approximated by the  equation~\citep{2006MNRAS.367..259H}
\begin{equation}\label{eq:color_temperature}
    T_\alpha\approx \frac{T_{\rm gas}}{1+0.41\left(\frac{1}{T_s}-\frac{1}{T_{\rm gas}}\right)},
\end{equation}
in terms of the spin temperature and the gas temperature. The collisional coupling coefficient  is given by
    \begin{equation}
        \begin{aligned}
        x_{\rm coll}&=x_{\rm coll}^{HH}+x_{\rm coll}^{eH}+x_{\rm coll}^{pH}\cr
        &=\frac{T_\star}{A_{10}T_{\rm CMB}}\left[k_{1-0}^{HH}(T_{\rm gas})n_{\rm H}+k_{1-0}^{eH}(T_{\rm gas})n_e+k_{1-0}^{pH}(T_{\rm gas})n_p\right],
        \end{aligned}
        \label{xcoll}
    \end{equation}
    where we have included the effects of collisions between two hydrogen atoms, and collisions between hydrogen and either an electron or a proton \cite{2012RPPh...75h6901P}. Here, $k_{1-0}^{ij}$ is the scattering rate between the species $i$ and $j$, and $n_i$ is the number density of species $i$ (with values that can be found {\it e.g}. in \citep{2019cosm.book.....M} and \citep{2006PhR...433..181F}).  
    
Finally, the gas temperature $T_{\rm gas}$ is in the simplest scenario determined by the rate of energy injection by the first massive stars, which also affects the number of ionized hydrogen in the IGM (parametrized by the electron fraction $x_e\equiv n_e/(n_e+n_{\rm H})$, with $n_e$ and $n_{\rm H}$ being the electron and neutral hydrogen densities respectively). Their evolution equations as a function of $x\equiv-\ln(1+z)$, $[dT_{\rm gas}/dx]_0$ and $[dx_e/dx]_0$, are included in  Appendix \ref{sec:app_std_evolution}. On the other hand, a population of DSs would also inject energy into the surrounding medium if annihilations into Standard Model particles occur. These inject energy in the IGM which will be absorbed mainly through three channels: IGM heating (heat), hydrogen ionization from the ground state (HIion) and excitation to the $n=2$ state (Ly$\alpha$). Due to these new effects, the evolution equations of  $T_{\rm gas}$ and $x_e$ are modified into~\cite{2007MNRAS.376.1680P,2018PhRvD..98d3006C,2018PhRvL.121a1103D,2020PhRvD.102j3005C,2021JCAP...05..051C}:
\begin{align}\label{eq:IGM_evol_mod_x}
\frac{dT_{\mathrm{gas}}}{dx}&=\left[\frac{dT_{\mathrm{gas}}}{dx}\right]_0+\frac{2}{3 n_{\rm H}(x)(1+f_{\rm He}+x_e)H(x)}\left[\frac{dE}{dVdt}\right]_{\rm dep,heat}, \nonumber \\
\frac{d x_{\mathrm{e}}}{dx}&=\left[\frac{dx_e}{dx}\right]_0+\frac{1}{n_{\rm H}(x)H(x)E_i}\left[\frac{dE}{dVdt}\right]_{\rm dep,HIion}+\frac{1-C_P}{n_{\rm H}(x)H(x)E_\alpha}\left[\frac{dE}{dVdt}\right]_{\rm dep,Ly\alpha},
\end{align}
where $E_i=13.6\ \rm eV$ is the hydrogen ionization energy, $E_{\alpha}= 3E_i/4=10.2\ \rm eV$ is the Ly$\alpha$ energy, $f_{\rm He}=0.245$ is the number fraction of helium, and  $C_P$ is the so-called Peeble's  factor, which has a complicated dependence on the gas temperature and the electron fraction that can be found in \cite{1968ApJ...153....1P,1995ApJ...455....7M,1995astro.ph..8126H}.

Finally $\left[\frac{dE}{dVdt}\right]_{\rm dep,c}$, $\mathrm{c}\in\{\mathrm{HIion},\,\mathrm{Ly}\alpha,\,\mathrm{heat}\}$, represent the energy deposited  into the IGM per comoving unit and time in the form of ionization (HIion), excitation (Ly$\alpha$) or heating (heat). We calculate them in the SSCK approximation~\citep{1982ApJS...48...95S,2004PhRvD..70d3502C}, which assumes that a fraction $f_{\rm eff}$ of the energy injected at some redshift is immediately transferred to the IGM, and that it is distributed among ionization, excitation and heating as~\citep{1982ApJS...48...95S,2004PhRvD..70d3502C}, 
\begin{equation}\label{eq:SSCK_fracs}
    f_{\rm HIion}=f_{\rm Ly\alpha}=f_{\rm eff}\frac{1-x_e}{3},\hspace{0.6cm}f_{\rm heat}=f_{\rm eff}\frac{1+2x_e}{3},    
\end{equation}
so that
\begin{equation}\label{eq:deposition_energy}
    \left[\frac{dE}{dVdt}\right]_{\rm dep,c}=f_{\rm c}\left(\frac{dE}{dVdt}\right)_{\rm inj}.
\end{equation}
In our scenario, DM annihilations produce Standard Model photons in the MeV–GeV energy range, as expected for dark electron masses in the GeV range. Such high-energy photons have a much smaller optical depth at $z\sim10$, and a large fraction will free-stream without depositing energy into the intergalactic medium. Therefore, the effective energy deposition fraction $f_{\rm eff}$ is significantly below unity. Based on the analysis in Ref.~\cite{Slatyer:2012yq} (see their Fig. 1), we adopt $f_{\rm eff}=0.001$ for our calculations. This choice accounts for the reduced efficiency of ionization and heating from high-energy photons.

\begin{figure}[t!]
    \centering
    \begin{subfigure}{0.45\textwidth}
        \includegraphics[width=\linewidth]{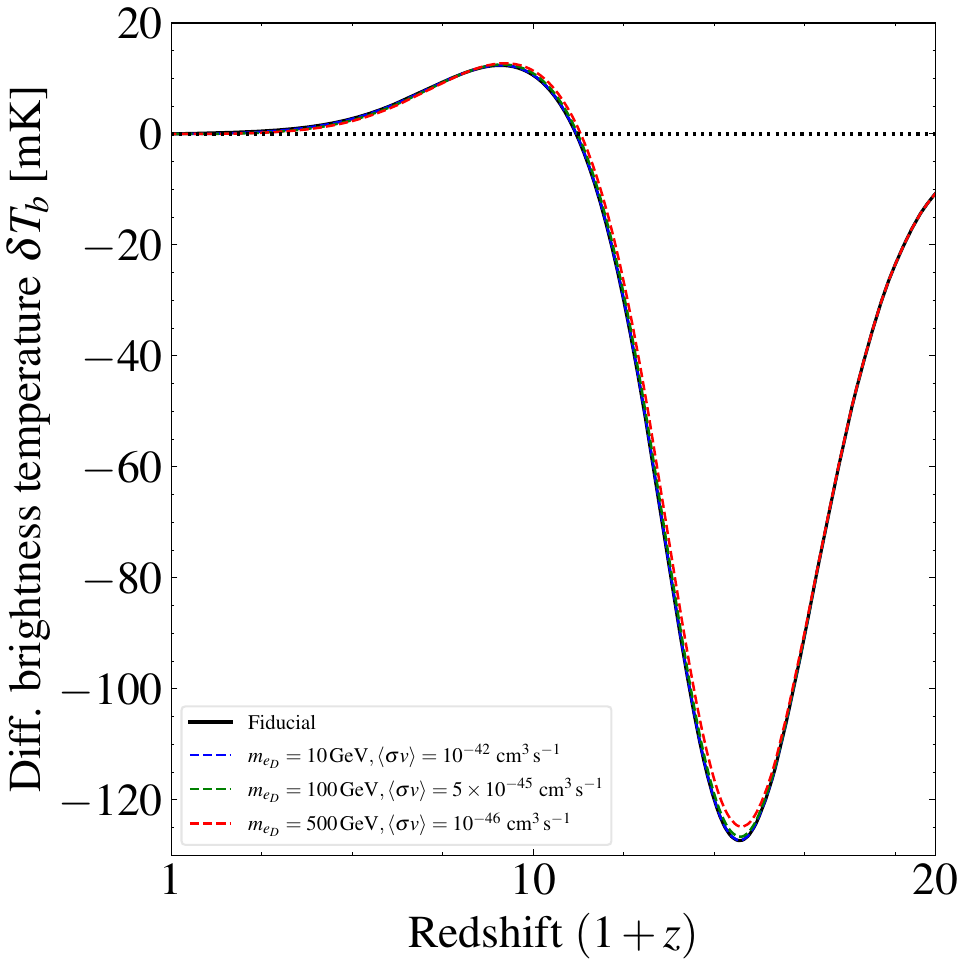}
    \end{subfigure}
    \begin{subfigure}{0.45\textwidth}
        \includegraphics[width=\linewidth]{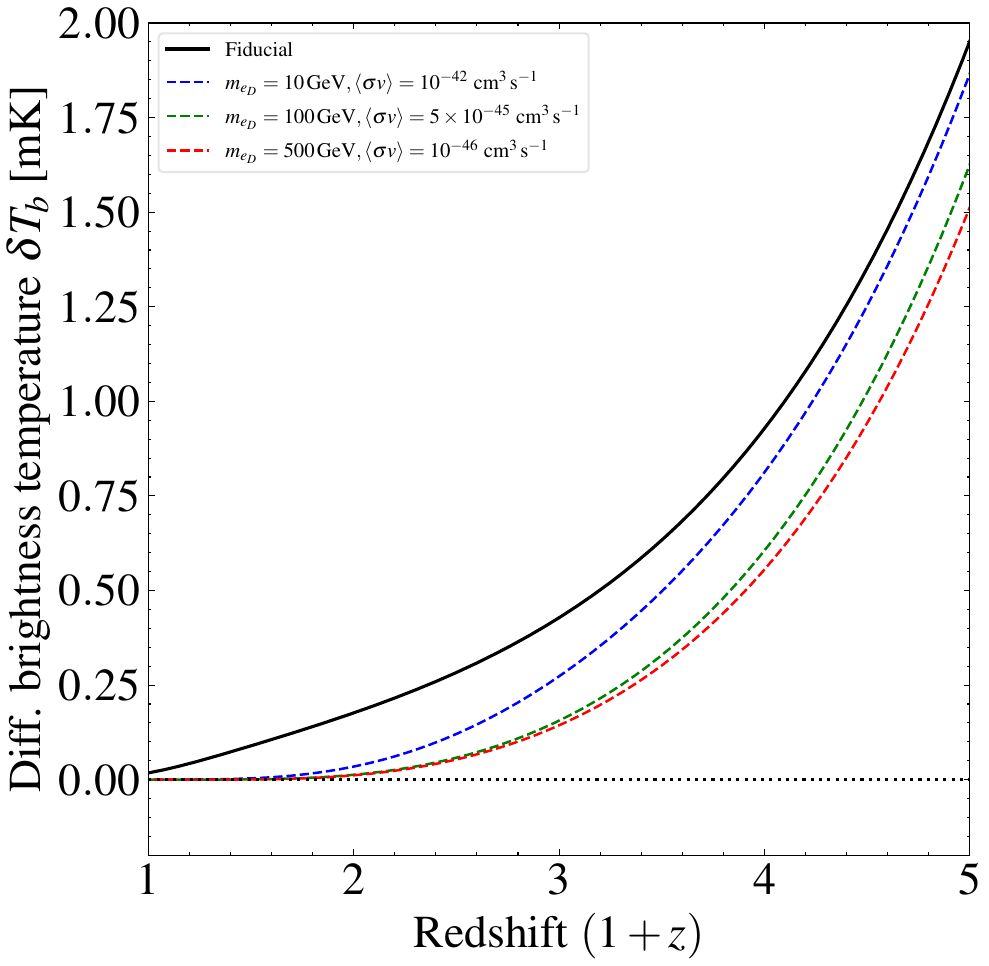}
    \end{subfigure}

    \begin{subfigure}{0.45\textwidth}
        \includegraphics[width=\linewidth]{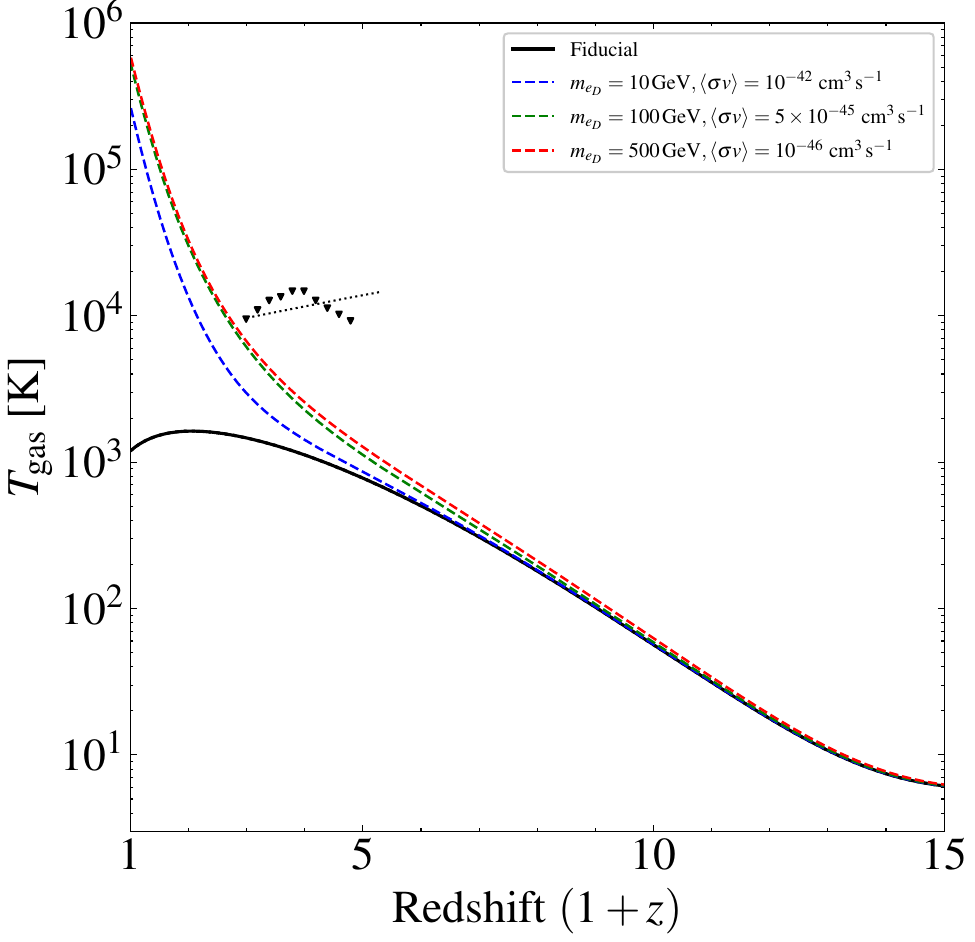}
    \end{subfigure}
    \begin{subfigure}{0.45\textwidth}
        \includegraphics[width=\linewidth]{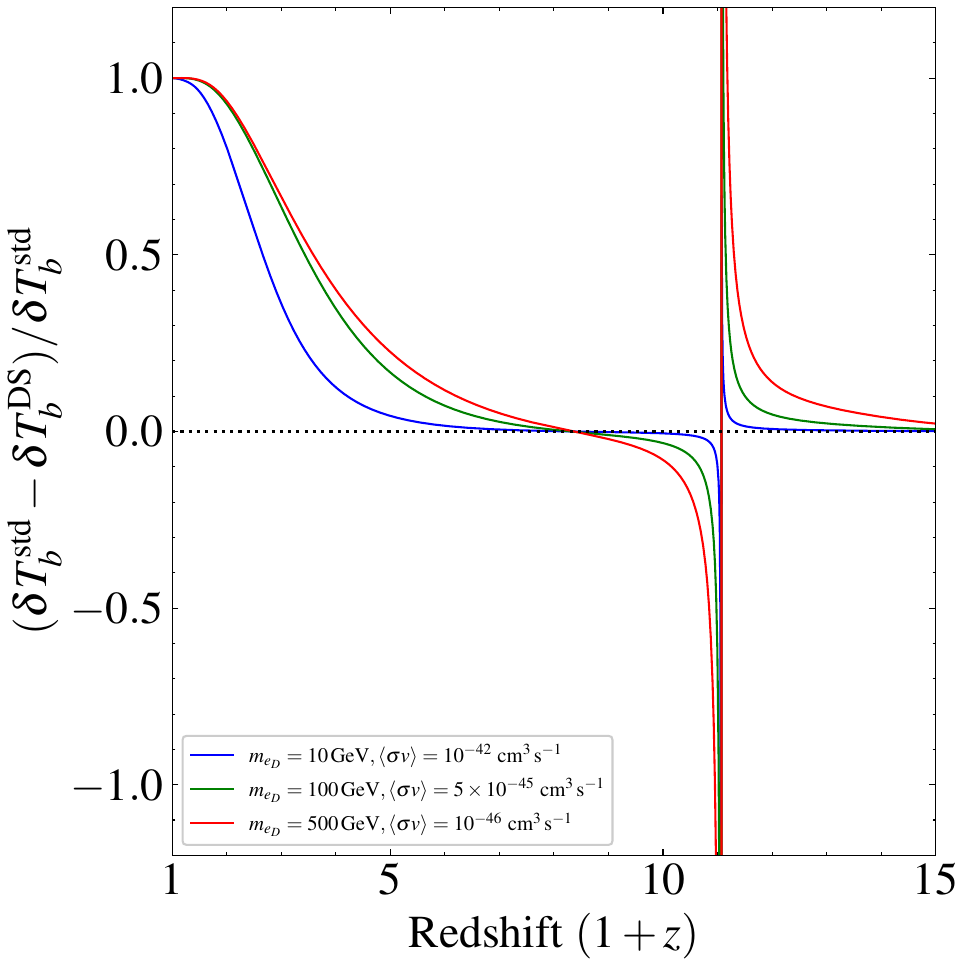}
    \end{subfigure}

    \caption{Top panels: Brightness temperature as a function of redshift. Bottom left panel: Gas temperature as a function of redshift. Bottom right panel: fractional comparison between the fiducial and the DS scenarios. In all panels, the black curve corresponds to the standard fiducial scenario, without inclusion of DSs. In all panels we fix $\alpha_D=0.1$ and $m_{\gamma_D}=100~\rm eV$.}
    \label{fig:signals}
\end{figure}

In Fig.~\ref{fig:signals}, we show the evolution of the brightness temperature (top panels), and gas temperature (bottom left panel) with the redshift for our benchmark dark electron masses (colored curves). The solid black line shows the expectation for these quantities in the standard fiducial scenario, where only the first stars inject energy. We take annihilation cross-sections of 
$10^{-42}$, $5\times10^{-45}$ and  $10^{-46}~\mathrm{cm}^3/\mathrm{s}$ for $m_{e_D}=$ 10, 100 and 500 GeV respectively. Furthermore, in the bottom left panel we present, as scatter points, measurements of the temperature of the universe up to redshifts of $z\sim4$ as performed by Ref.~\cite{Gaikwad:2020eip}. Furthermore, the black dotted line corresponds to previous constraints obtained by Ref.~\cite{Schaye:1999vr}.

We also present in the bottom right panel the fractional comparison between the fiducial and the DS scenarios. In all panels, we adopted the density profiles shown in the lower panel of Fig.~\ref{fig:mass_vs_r}, corresponding to a DM scenario with $m_{\gamma_D}=100$ eV and $\alpha_D=0.1$. Concretely, we have adopted the density profiles shown in the lower-right plot of Fig.~\ref{fig:mass_vs_r},  which amount to a DS luminosity in photons
\begin{align}
	L_{\rm ann}
	&\simeq  (0.00036,2.864,1.75\times10^3)\, L_\odot\, f^\gamma_{\rm ann}\left(\frac{\langle\sigma v\rangle_{\rm ann}}{10^{-46}\ \mathrm{cm}^3\, \mathrm{s}^{-1}}\right),
\end{align}
for $m_{e_D}=(10,100,500)$ GeV respectively. Moreover, we have assumed for simplicity that all the DSs are identical and that they constitute a fraction $f=0.05$ of the total DM of the Universe. 

Regarding the bottom left panel with the evolution of the gas temperature, we note that, given that the main avenue of energy injection goes into heating the IGM, the gas can reach substantial temperatures at low redshift. Nevertheless, we must make sure that the matter temperature does not exceed the measurements given by the scattered points. For our choice of DS benchmarks, we see that we avoid this constraint. However, this has consequences in the strength of the modification of the differential brightness temperature. We observe that the heating of the gas due to the energy injection by DSs increases the minimum brightness temperature by a few percent and shifts it to higher redshifts, as is made apparent by the bottom right panel. The effect is more accentuated as the cross-section increases (for fixed DM mass). We can also note that the maximum of the signal is also slightly increased and shifted to larger redshift. The most important variation of the signal occurs at smaller redshift. As the DS scenarios drive reionization faster than the fiducial case, the brightness temperature disappears earlier than expected, leading to a difference that can even be double of what is expected in the fiducial case.

At this point, we would like to acknowledge an important caveat. The modification of the 21-cm signal introduced by dark stars may be degenerate with variations in the uncertain astrophysical parameters of the standard model. For instance, a stronger X-ray background from early binaries or a different star formation efficiency could also produce additional heating. While we have shown that DSs are likely to form before baryonic stars, their imprints on the signal happen at the redshifts where the first astrophysical sources also become active. A rigorous statistical analysis will be necessary to break the degeneracies between the signature of DS annihilation in compact objects and the uncertain standard parameters.

One might wonder how a small fraction of DM, with such a weak annihilation cross-section can impact the 21-cm signal. To gain some intuition, let us perform a rough estimate comparing the heating contributions from Population II and III stars and DSs. Baryonic matter constitutes about $\frac15$ of the total DM content, and approximately half of that is expected to form stars. Population III stars are typically supermassive, with masses ranging from several tens up to $\sim 300 M_{\odot}$. For simplicity, we consider a representative star of mass $100 M_{\odot}$. These stars follow a luminosity scaling of $L \sim 10^4 L_{\odot} (M/M_{\odot}) $, yielding $L \sim 10^6 L_{\odot}$ for our chosen mass. In contrast, DSs in our scenario form from a 10\% subcomponent of DM, with about 50\% of that forming DSs, mirroring the star formation efficiency assumed for baryons. The typical mass of a DS is $\sim 10 M_{\odot}$, implying that DSs are roughly 200 times more numerous than ordinary stars due to their lower mass. Despite the small annihilation cross-section, the compactness of DSs leads to extremely high DM densities in their cores. Combined with the fact that essentially all DM mass is converted into radiation, this results in DS luminosities around $10^3 L_{\odot}$.

Though DSs are less massive, their high number density and significant luminosity mean that their collective energy output can surpass that of Population II and III stars by roughly a factor of 20. While this is a ballpark estimate, it highlights how even a small DM component can have an impact on the 21-cm signal through DS formation.

\section{Conclusions}
\label{sec:conclusions}

A small but strongly interacting component of asymmetric DM can undergo gravitational collapse to form compact DSs. The extreme dark matter densities achieved in these objects can lead to significant annihilation into Standard Model particles, even for annihilation cross-sections that are highly suppressed and currently undetectable by other means.

In this work, we have demonstrated that under specific conditions, a population of DSs can form concurrently with or prior to the first baryonic stars. The subsequent annihilations within these DSs inject energy into the intergalactic medium, predominantly in the form of heating. We have shown that this energy injection leaves an imprint on the global 21-cm signal from the cosmic dawn era, which deviates from the predictions of standard astrophysical models.

Contrary to scenarios that primarily enhance ionization, the dominant effect of DS annihilations is heating of the IGM. This manifests in the 21-cm signal not as a dramatically deeper absorption feature, but through other modifications. A slight elevation (by a few percent) and a shift to higher redshift of the minimum brightness temperature, a corresponding shift of the emission peak to higher redshift and an earlier end of reionization, resulting in a more rapid decline of the 21-cm signal at lower redshifts.

Several directions can improve this work. A more realistic DS formation model that considers baryonic feedback and the nonlinear process of the final stages of the collapse is needed. For example, studying how the collapse of the strongly interacting DM component evolves via numerical simulations will give a more accurate estimate of redshift at which DSs start affecting the 21-cm spectrum by injecting energy. Another direction will be to study the effect not for the case where DSs have the same mass, but for a more realistic scenario with a distribution of masses, as it happens with ordinary stars.

\acknowledgments

We thank Miguel Escudero Abenza and Paolo Panci for useful discussions. for The work of BBK and AI was supported by the Deutsche Forschungsgemeinschaft (DFG, German Research Foundation) under Germany’s Excellence Strategy - EXC-2094 - 390783311 and under the Collaborative Research Center SFB1258 grant - SFB-1258 - 283604770. BBK is supported by IBS under the project code IBS-R018-D3.

\clearpage

\appendix
\section{Calculation of the Ly$\alpha$ coupling to the spin temperature}\label{sec:app_lyalpha}
In this appendix we summarize the computation of the Ly$\alpha$ coupling to the spin temperature given in Eq. (\ref{eq:Ly_alpha_coup}). To this end, it is necessary to calculate the correction term $S_\alpha$ and the Ly$\alpha$ background intensity $J_\alpha$, which is detailed in what follows.
\subsection{Radiative transfer factor}
The term $S_{\alpha}$ is a correction factor which takes into account complicated radiative transfer effects in the intensity near the line center and causes suppression of the radiation spectrum. We adopt the approximation given by \cite{2006MNRAS.372.1093F}, which is accurate to a few percent for $T_{\rm gas}\geq1\ \rm K$:
\begin{equation}
    S_{\alpha}\approx\mathrm{exp}\left[-1.12\eta'\left(\frac{3a}{2\pi\gamma'}\right)^{\frac13}\right].
\end{equation}
In this expression $a=\frac{A_{21}}{4\pi\Delta\nu_D}$ is the Voigt parameter with $A_{21}=6.25\times10^8\ \mathrm{s}^{-1}$ the spontaneous emission coefficient for the Ly$\alpha$ transition. The term $\Delta\nu_D=\sqrt{\frac{2 T_{\rm gas}}{m_{\rm H}}}\nu_\alpha$ is the Doppler width with $\nu_{\alpha}=2.47\times10^{15}\,\mathrm{Hz}$ the central frequency of the Ly$\alpha$ transition and $m_{\rm H}$ the hydrogen mass \cite{2006MNRAS.372.1093F}. 
Furthermore $\gamma'=\gamma\left(1+\frac{0.4\ \mathrm{K}}{T_{\rm gas}}\right)^{-1}$ is the correction due to spin exchange to the Sobolev parameter $\gamma=\tau_{\rm GP}^{-1}$, which is written in terms of the inverse of the Gunn-Peterson optical depth \cite{2006PhR...433..181F}. This optical depth is given by $\tau_{\rm GP}=\frac{\pi e^2 n_{\rm H}(x) f_{12}}{H(x)m_e \nu_\alpha}$ where $f_{12}=0.416$ is the oscillator strength for the Ly$\alpha$ transition \cite{2004ApJ...602....1C}. Additionally, $\eta=\frac{2\pi\nu_\alpha^2}{m_p  \Delta\nu_D}$ and
\begin{equation}
    \eta^\prime=\eta\left(\frac{1+\frac{0.4\ \rm K}{T_s}}{1+\frac{0.4\ \rm K}{T_{\rm gas}}}\right)-(x+x_0)^{-1},
\end{equation}
which for $T_{\rm gas}>>0.4\ \rm K$ can be approximated as $\eta^\prime\approx\eta-(x+x_0)^{-1}$. Here $x_0=\frac{\nu_\alpha}{\Delta\nu_D},x=\frac{\nu-\nu_\alpha}{\Delta\nu_D}$: Close to the line center $|x|<<x_0$, so we finally approximate $\eta^\prime\approx\eta-x_0^{-1}$.

\subsection{Ly$\alpha$ background intensity}
The average Ly$\alpha$ background $J_\alpha$ consists of two components \citep{2011MNRAS.411..955M}: the direct emission of photons between the Ly$\alpha$ energy $E_{\alpha}$, and the Lyman limit $E_i$, which is usually produced from Population II and III stars ($J_{\alpha,\star}$) and X-ray excitation of HI ($J_{\alpha,X}$). The total background is then
\begin{equation}
    J_\alpha=J_{\alpha,\star}+J_{\alpha,X},
\end{equation}
and in the following Subsections we provide details on the calculation of both terms.
\subsubsection{Direct emission}\label{app:direct}
In the case of $J_{\alpha,\star}$, it was pointed out by \cite{2006MNRAS.367.1057P} that not only the contribution to the Ly$\alpha$ background of photons emitted with an energy $E_\alpha$ was important, but rather photons emitted in the whole Lyman series played an important role, as they are redshifted into the Lyman resonance. As such, the background from photons that redshift into the Lyman resonance is calculated as a sum over the Lyman series ($\nu_n=R_H\left(1-\frac{1}{n^2}\right)$ with $R_H=1.0973\times10^7\ \mathrm{m}^{-1}$ the Rydberg constant) and is given by
\begin{equation}\label{eq:Lyalpha_back_direct}
    J_{\alpha,\star}(z)=\sum_{n=2}^{n_{\rm max}}{J_{\alpha}^{(n)}(z)}=\sum_{n=2}^{n_{\rm max}}{f_{\rm recycle}(n)\int_{z}^{z_{\rm max}(n)}{dz^\prime\frac{(1+z)^2}{4\pi}\frac{1}{H(z^\prime)}\hat{\epsilon}_\star(\nu_n^\prime,z^\prime)}},
\end{equation}
where $\nu_n^\prime=\nu_n\frac{1+z^\prime}{1+z}$ is the frequency at redshift $z^\prime$ that redshifts into the resonance at redshift $z$. The largest redshift from which a photon can redshift into the resonance is $1+z_{\max}(n)=(1+z)\frac{1-(n+1)^{-2}}{1-n^{-2}}$ \cite{2006MNRAS.371..867F}. The factor $f_{\rm recycle}$ is the probability that a Ly$n$ photon will generate a Ly$\alpha$ photon and are provided in table \ref{tab:recycling_fractions} with values from \cite{2006MNRAS.367.1057P}. The sum is truncated at $n_{\rm max}\approx23$, to exclude levels for which the photons reach the HII region of a typical (isolated) galaxy, as only neutral hydrogen contributes to 21-cm absorption.
\begin{table}[th!]
\centering \small \addtolength{\tabcolsep}{2pt}
\caption{Recycling fractions.}
\label{tab:recycling_fractions}
\begin{tabular}{c c c c c c c c}
 \hline \hline
\scriptsize{$n$} & \scriptsize{$f_{\rm recycle}(n)$} & \scriptsize{$n$} & \scriptsize{$f_{\rm recycle}(n)$} & \scriptsize{$n$} & \scriptsize{$f_{\rm recycle}(n)$} & \scriptsize{$n$} & \scriptsize{$f_{\rm recycle}(n)$} \\
\hline
$2$  & $1$       & $10$ & $0.3476$  & $18$ & $0.3561$  & $26$ & $0.3584$  \\
$3$  & $0$       & $11$ & $0.3496$  & $19$ & $0.3565$  & $27$ & $0.3586$  \\
$4$  & $0.2609$  & $12$ & $0.3512$  & $20$ & $0.3569$  & $28$ & $0.3587$  \\
$5$  & $0.3078$  & $13$ & $0.3524$  & $21$ & $0.3572$  & $29$ & $0.3589$  \\
$6$  & $0.3259$  & $14$ & $0.3535$  & $22$ & $0.3575$  & $30$ & $0.3590$  \\
$7$  & $0.3353$  & $15$ & $0.3543$  & $23$ & $0.3578$  &      &           \\
$8$  & $0.3410$  & $16$ & $0.3550$  & $24$ & $0.3580$  &      &           \\
$9$  & $0.3448$  & $17$ & $0.3556$  & $25$ & $0.3582$  &      &           \\
\hline \hline
\end{tabular}
\end{table}

\noindent The term
\begin{equation}
    \hat{\epsilon}_\star(\nu,z)=\hat{\epsilon}_\star(\nu)f_\star n_{m,0}\frac{df_{\rm coll}}{dt},
\end{equation}
is the comoving emissivity at frequency $\nu$ and it models the energy that is deposited by a given source. Here, $n_{m,0}$ is the  matter number density today that produces the sources that emit photons (in the case of ordinary stars, it is the baryon number density; for DSs it's the DM number density), $f_{\star}$ is the star formation efficiency (in the case of DSs it is $f_{\star,\rm DS}$) and $\hat{\epsilon}_\star(\nu)$ is the spectral energy distribution of the sources, which is the number of photons per frequency emitted at $\nu$ per baryon (DM particle) in stars (DSs).

\noindent The term 
\begin{equation}\small
    f_{\rm coll}=\rho_{\rm m}(z)^{-1}\int_{M_{\rm min}}^{M_{\rm max}}M\frac{dn}{dM}dM=\mathrm{erfc}\left(\frac{\delta_c(z)}{\sqrt{2}\sigma(M_{\rm min}(z))}\right)-\mathrm{erfc}\left(\frac{\delta_c(z)}{\sqrt{2}\sigma(M_{\rm max}(z))}\right),
\end{equation}
is the fraction of matter collapsed into halos with $M_{\rm min}<M<M_{\rm max}$~\cite{2014MNRAS.443.1211M}, where in the last equality we have particularized to the Press \& Schechter model. For ordinary stars, $M_{\rm max}$ is taken to infinity. Furthermore the minimum halo mass corresponds to a halo with virial temperature
$T_{\rm vir}=10^4\ \rm K$. This is the threshold for atomic cooling which allows the baryonic gas to fragment and form stars \cite{2001PhR...349..125B,2006MNRAS.371..867F}. The virial temperature is \citep{2013fgu..book.....L}
\begin{equation}\label{eq:T_virial}
    T_{\rm vir}=1.98\times10^{4}\ \mathrm{K}\left(\frac{1+z}{10}\right)h^{\frac23}\left(\frac{\mu}{0.6}\right)\left[\frac{\Omega_{m,0}}{\Omega_{m}(z)}\frac{\Delta_c(z)}{18\pi^2}\right]^{\frac13}\left(\frac{M_{\rm halo}}{10^8M_\odot}\right)^{\frac23},
\end{equation}
where $\mu$ is the mean atomic weight of the neutral primordial gas in units
of the proton mass (kept fixed here) and $\Delta_c(z)$ is the virial overdensity \citep{1998ApJ...495...80B,2015JCAP...06..005M}
\begin{equation}
    \Delta_c(z)=18\pi^2+82\left(\Omega_m(z)-1\right)-39\left(\Omega_m(z)-1\right)^2,
\end{equation}
where $\Omega_m(z)=\Omega_{m,0}(1+z)^3/(H(z)/H_0)^2$. In the case of DSs, $f_{\rm coll}$ is readily found from Eq.~\eqref{eq:f_coll_DS} with the same considerations from the encompassing Section.

For Population III and II stars, there is an empirical function $\hat{\epsilon}_\star(\nu)=\hat{\epsilon}_{\star,III}(\nu)+\hat{\epsilon}_{\star,II}(\nu)$ ~\cite{2005ApJ...626....1B}, with
\begin{equation}\label{eq:sdf_III}
    \hat{\epsilon}_{\star,III}(\nu)=
    \begin{cases}
        \left(6.291\times10^{-12}\ \frac{\rm photons}{\rm baryons\times Hz}\right)\left(\frac{\nu}{\nu_i}\right)^{0.29},\hspace{0.5cm}\nu_{\alpha}<\nu<\nu_{\beta},\cr
        \left(5.859\times10^{-12}\ \frac{\rm photons}{\rm baryons\times Hz}\right)\left(\frac{\nu}{\nu_i}\right)^{0.29},\hspace{0.5cm}\nu_{\beta}<\nu<\nu_{i},
    \end{cases}
\end{equation}
and
\begin{equation}\label{eq:sdf_II}
    \hat{\epsilon}_{\star,II}(\nu)=
    \begin{cases}
        \left(1.217\times10^{-11}\ \frac{\rm photons}{\rm baryons\times Hz}\right)\left(\frac{\nu}{\nu_i}\right)^{-0.86},\hspace{0.5cm}\nu_{\alpha}<\nu<\nu_{\beta},\cr
        \left(8.124\times10^{-12}\ \frac{\rm photons}{\rm baryons\times Hz}\right)\left(\frac{\nu}{\nu_i}\right)^{-0.86},\hspace{0.5cm}\nu_{\beta}<\nu<\nu_{i},
    \end{cases}
\end{equation}
corresponding to the spectral distribution functions for Population III and II stars, respectively. The frequencies are $\nu_\alpha=2.47\times10^{15}\ \rm Hz$ (equivalent to the Ly$\alpha$ energy $E_\alpha$), $\nu_\beta=2.92\times10^{15}\ \rm Hz$ (equivalent to the Ly$\beta$ energy $E_\beta=12.08\ \rm eV$) and $\nu_i=3.29\times10^{15}\ \rm Hz$ (the Lyman-limit, equivalent to the ionization energy $E_i$). The function (\ref{eq:sdf_III}) is normalized such that $2,670\ \frac{\rm photons}{\rm baryon}$ are emitted between Ly$\alpha$ and Ly$\beta$ and $4,800\ \frac{\rm photons}{\rm baryon}$ are emitted between Ly$\alpha$ and the Lyman-limit, in accordance with \cite{2001ApJ...552..464B}. On the other hand, the function (\ref{eq:sdf_II}) is normalized such that $6,520\ \frac{\rm photons}{\rm baryon}$ are emitted between Ly$\alpha$ and Ly$\beta$ and $9,690\ \frac{\rm photons}{\rm baryon}$ are emitted between Ly$\alpha$ and the Lyman-limit, in accordance with \cite{1999ApJS..123....3L}.

In our case of interest i.e., DSs, the spectral distribution function is
\begin{align}
    \small
    \hat{\epsilon}_\star(\nu)&=\frac{L_{\rm ann}\tau_{\rm DS}}{N_{\rm DS}m_{e_D}}\,\delta\left(\nu-\frac{m_{e_D}}{2\pi}\right)\cr
    &=\left(5.73\times10^{-5}\ \frac{\rm photons}{\text{DM particles}}\right)f_{\rm ann}^\gamma\lambda^{-1}\left(\frac{m_{e_D}}{100\ \rm MeV}\right)^{3}\cr
    &\times\left(\frac{\langle\sigma v\rangle_{\rm ann}}{1.2\times10^{-58}\ \mathrm{cm}^3\ \mathrm{s}^{-1}}\right)\left(\frac{\tau_{\rm DS}}{10\ \rm Gyr}\right)\left(\frac{J_\star}{10^{-5}}\right)\delta\left(\nu-2.428\times10^{22}\ \rm Hz\right),
\end{align}
where $N_{\rm DS}=\frac{M_{\rm DS}}{m_{e_D}}$ is the number of DM particles in the DS, $\tau_{\rm DS}$ is the DS's lifetime and the $\delta$ function imposes that the energy is released monochromatically at the annihilation energy $m_{e_D}$. However, we notice that given the difference over orders of magnitude between the DM mass and the energy of the Lyman series $\hat{\epsilon}_\star(\nu_n)=0$ for all $n$. As such, DSs do not contribute to the Ly$\alpha$ background according to Eq. (\ref{eq:Lyalpha_back_direct}).
\subsubsection{X-ray excitation}
X-rays deposit energy in the IGM by ionizing hydrogen and helium; the emitted electron then distributes its energy via i) Coulomb collisions with thermal electrons, ii) collisional ionization, which produces more secondary electrons and iii) collisional excitation of helium, which produces a photon capable of ionizing hydrogen. The contribution to the Ly$\alpha$ background due to these processes can be related to the X-ray heating rate $\epsilon_X$ as \cite{2007MNRAS.376.1680P}

\begin{equation}
    J_{\alpha,X}=\frac{1}{8\pi^2}\frac{\epsilon_{X,\alpha}}{\nu_\alpha}\frac{1}{H\nu_\alpha},
\end{equation}
with the conversion rate of X-ray energy into Ly$\alpha$ photons per unit comoving volume given by $\epsilon_{X,\alpha}=\epsilon_{X,\rm tot} \frac{f_{\rm X,exc}}{f_{\rm X,heat}}p_\alpha$, where $p_\alpha\approx0.79$ is the fraction of excitation energy that goes into Ly$\alpha$ photons.  The fraction of X-ray energy going to heating $f_{\rm X,heat}$, ionization $f_{\rm X,ion}$ and excitation $f_{\rm X,exc}$ has been treated by e.g \cite{1985ApJ...298..268S}, which fitted the results (for primary electron energy $E\geq100\ \rm eV$) as
\begin{equation}\label{eq:X_ray_fractions}
    \begin{cases}
        f_{X,\rm heat}=0.9971\left[1-(1-x_{\rm e}^{0.2663})^{1.3163}\right],\cr
        f_{X,\rm ion}=0.3908\left(1-x_{\rm e}^{0.4092}\right)^{1.7592},\cr
        f_{X,\rm exc}=0.4766\left(1-x_{\rm e}^{0.2735}\right)^{1.5221}.
    \end{cases}
\end{equation}
The term $\epsilon_{X,\rm tot}=\epsilon_{X,\rm std}+\epsilon_{X,\rm DS}$ is the total heating rate and takes into account both the standard and the DS contributions. The heating rate in the standard (no DM) model can be estimated as \cite{2006MNRAS.371..867F}
\begin{equation}\label{eq:epsilon_X}
    \epsilon_{X,\rm std}=\left(\frac{3 n_{\rm H} (1+f_{\rm He}+x_e) H(z)}{2}\right)\left[10^3\ \mathrm K\, f_X\left(\frac{f_\star}{0.1}\frac{f_{X,\rm heat}}{0.2}\frac{df_{\rm coll}/dz}{0.01}\frac{1+z}{10}\right)\right],
\end{equation}
where $f_X$ is the X-ray luminosity per star formation rate relative to the local value (usually taken as a free parameter of the order $\sim0.5$), $f_\star$ is the star formation efficiency and $df_{\rm coll}/dz$ is, as previously mentioned, the evolution of the collapsed fraction.

In the case of DSs, the heating rate can be quickly obtained from Eq. (\ref{eq:deposition_energy}) as 
\begin{equation}
    \epsilon_{X,\rm DS}=\left[\frac{dE}{dVdt}\right]_{\rm dep,heat}.
\end{equation}
\section{Standard evolution equations}\label{sec:app_std_evolution}
As mentioned previously, the bracketed terms in Eq.~(\ref{eq:IGM_evol_mod_x}) with the $0$ subscript denote the standard evolution equations with no DS contribution (see Refs. \citep{2006PhR...433..181F,2012RPPh...75h6901P,2023JApA...44...10B} for comprehensive reviews) and take the form
\begin{align}\label{eq:IGM_evol}
    \left[\frac{dx_e}{dt}\right]_0&=C_P\left[\beta(T_{\rm gas})(1-x_e)-n_{\rm H}\alpha^{(2)}(T_{\rm gas}) x_e^2\right]+\zeta\left(\frac{df_{\rm coll}}{dt}\right),\cr
    \left[\frac{dT_{\mathrm{gas}}}{dt}\right]_0&=-2H(z) T_{\rm gas}+\gamma_c(T_{\rm CMB}-T_{\rm gas})+\left[\frac{d T_{\rm gas}}{dt}\right]_{0. \rm heat},
\end{align}
where $T_{\rm CMB}=2.7255\ \mathrm{K}(1+z)$ is the CMB temperature, $\beta(T)=\alpha^{(2)}(T)\left(\frac{m_{\mathrm{e}} T}{2\pi}\right)^{\frac32}\mathrm{e}^{-\frac{E_i}{T}}$ and $\gamma_c$ is the coefficient defined as
\begin{equation}
    \gamma_c\equiv\frac{8\sigma_T a_r T_{\rm CMB}^4}{3m_e}\left(\frac{x_e}{1+f_{\rm He}+x_e}\right),
\end{equation}
with $a_r=7.5657\times10^{-16}\,\mathrm{J}\,\mathrm{m}^{-3}\,\mathrm{K}^{-4}$, the radiation constant.

The parameter $\zeta=A_{\rm He}f_\star f_{\rm esc}N_{\rm ion}$ is the ionization efficiency parameter~\cite{2006MNRAS.371..867F,2007MNRAS.376.1680P} with $N_{\rm ion}$ the number of ionizing photons per baryon produced in stars, $f_{\rm esc}$ the fraction of ionizing photons that escape their host galaxy and $A_{\rm He}=4/(4-3f_{\rm He})$ is a correction factor to convert the number of ionizing photons per baryon in stars to the fraction of ionized hydrogen. Aside from $A_
{\rm He}$, the parameters $f_{\rm esc}$ and $N_{\rm ion}$ are highly uncertain. In this case, we follow the fiducial case of Ref.~\cite{2006MNRAS.371..867F} and consider a ionization efficiency of $\zeta=73.2$ with $f_{\rm esc}=0.02$ and $N_{\rm ion} = 30,000$. This is the case that assumes Population III stars are the dominant source of ionizing radiation.

We also recast these equations as in Section~\ref{sec:signals} with the parameter $x\equiv-\ln(1+z)$ to obtain
\begin{align}\label{eq:IGM_evol_x}
    \left[\frac{dx_e}{dx}\right]_0&=\frac{C_P}{H(x)}\left[\beta(T_{\rm gas})(1-x_e)-n_{\rm H}\alpha^{(2)}(T_{\rm gas}) x_e^2\right]+\frac{\zeta}{H(x)}\left(\frac{df_{\rm coll}}{dt}\right),\cr
    \left[\frac{dT_{\mathrm{gas}}}{dx}\right]_0&=-2T_{\rm gas}+\frac{\gamma_c}{H(x)}(T_{\rm CMB}-T_{\rm gas})+\left[\frac{d T_{\rm gas}}{dx}\right]_{0. \rm heat}.
\end{align}
The heating term in the gas temperature equation is of the form
\begin{equation}
    \left[\frac{d T_{\rm gas}}{dt}\right]_{0. \rm heat}=\frac{2}{3}\sum\limits_{c}{\frac{\epsilon_c}{n_{\rm H}(1+f_{\rm He}+x_e)H(x)}},
\end{equation}
where the sum is over all possible contributions to the heating; namely X-rays and Ly$\alpha$ photons. It is important to note that in this expression, the X-ray component is only due to the standard (no DM) scenario, as all the DS contribution is already taken into account in the additional term of Eq. (\ref{eq:IGM_evol_mod_x}). For the X-ray contribution, the heating term can be easily found from the rate Eq. (\ref{eq:epsilon_X}). For the Ly$\alpha$ photons, the heating term is calculated according to \cite{2006PhR...433..181F}
\begin{equation}
    \frac{2}{3}\frac{\epsilon_\alpha}{n_{\rm H}(1+f_{\rm He}+x_e) H(z)}=\frac{16\pi^2}{3}\frac{\nu_\alpha J_{\alpha,\star} \Delta\nu_D}{n_{\rm H}(1+f_{\rm He}+x_e)}I_C,
\end{equation}
where
\begin{equation}
    I_C=\left(\frac{4}{\pi}\right)^{-\frac16}\pi^{\frac32}\left(\frac{a}{\gamma'}\right)^{\frac13}\beta\left[Ai^2(-\beta)+Bi^2(-\beta)\right],
\end{equation}
with $Ai(x)$ and $Bi(x)$ the Airy functions \cite{2006MNRAS.372.1093F}. The background intensity $J_{\alpha,\star}$ is calculated according to eqs. (\ref{eq:Lyalpha_back_direct},~\ref{eq:sdf_III},~\ref{eq:sdf_II}). Furthermore, we write $\beta=\eta'\left(\frac{4a}{\pi\gamma'}\right)^{\frac13}$, where all quantities involved are as  defined in appendix \ref{sec:app_lyalpha}.

\bibliographystyle{JHEP} % We choose the "plain" reference style
\bibliography{References} % Entries are in the References.bib file

\end{document}